\documentclass[12pt]{iopart}

\usepackage{iopams}  

\usepackage{graphicx}
\usepackage{epsf}
\usepackage{mathptmx}
\begin{document}

\title[On A 5-Dimensional Spinning Cosmic String] {On A 5-Dimensional Spinning Cosmic String }

\author{R J Slagter}

\address{Institute of Physics, University of Amsterdam and ASFYON
Pr Marielaan 16 1405EP Bussum, The Netherlands}
\ead{info@asfyon.nl}
\begin{abstract}
We present a numerical solution  of a stationary 5-dimensional spinning cosmic string in the Einstein-Yang-Mills (EYM) model,
where the extra bulk coordinate $\psi$ is periodic.
It turns out that when $g_{\psi\psi}$ approaches zero, i.e., a closed time-like curve (CTC) would appear, the solution becomes singular.
When a negative cosmological constant is incorporated in the model, the solution becomes regular everywhere with angle deficit$<2\pi$.
However, the cosmic string-like object has not all the desired asymptotic properties of the counterpart Abelian Nielsen-Olesen string.
When we use a two point boundary value routine with the correct cosmic string features far from the core, then again a negative $\Lambda$
results in an acceptable string-like solution.

We also investigated the possibility of a Gott space time  structure of the static 5D cosmic string. The matching condition yields no obstruction
for an effective angle deficit. Moreover, by considering the angular momentum in bulk space , no helical structure of time is
necessary. Two opposite moving 5D strings could, in contrast with the 4D case, fulfil the Gott condition.

\end{abstract}

\vspace{2pc}
\maketitle

\section{Introduction}
In recent years higher dimensional gravity is attracting much interest. One reason is the possibility that these higher
dimensions could become detectable at CERN.
The possibility that spacetime may have more than four dimensions is initiated by high energy physics
and inspired by D-brane ideology in string theory.
Our 4-dimensional spacetime (brane) is embedded in the 5-dimensional bulk. It is assumed that all the standard model
degrees of freedom reside on the brane, where as gravity can propagate into the bulk \cite{Ran}.
The effect of string theory on classical gravitational physics is investigated by the low-energy effective action.
If our 5 dimensional space time is obtained as an effective theory, the matter fields, for example the U(1) field, can exist
in the bulk.
In General Relativity(GR), gravitating non-Abelian gauge field, i.e., the Yang-Mills(YM) field, can be regarded as the most natural
generalization of Einstein-Maxwell(EM) theory. In particular, particle-like, soliton-like  and black hole solutions in the combined
Einstein-Yang-Mills(EYM) models, shed new light on the complex features of compact object in these models. See \cite{Vol} for an overview.
The reason for adding a cosmological constant to these models was inspired by the study of the so-called AdS/CFT correspondence \cite{Mal,Hos},
since the 5-dimensional Einstein gravity with cosmological constant gives a description of 4-dimensional conformal field theory
in large N limit. Moreover, in brane world scenarios, negative cosmological constant is naturally expected.

Gravitating cosmic strings  became of interest, when it was discovered that inflationairy cosmological models solved many shortcomings
in the standard model. Inflation is triggered by a Higgs field ($\Phi$) on the right hand side of the equations of Einstein. This complex scalar field,
with the "Mexican hat" potential, was also necessary in Ginzburg-Landau model as order parameter to explain the famous Meissner effect in
superconductivity. This Abelian Higgs model also yields the Nielsen-Olesen vortex solution in flat space as well as in
curved space time. This model contains besides the $\Phi$ field also a gauge field ($A_\mu$)  and is invariant under  U(1) global
phase transitions. The vacuum of this model is degenerated,
i.e., not invariant under this transformation. The result is a topological defect:
when $\Phi$ winds once around the vacuum manifold, one obtains a contradiction with single-value of $\Phi$.
So $\Phi$ rises to the top of the potential, and a lot of potential energy is stored in the scalar field configuration. They
form lines of trapped energy density: a cosmic string. One obtains a quantized magnetic flux which is $\frac{2\pi n}{e}$, with
n the winding number and $e$ the electric charge. Due to the fact that strings have stress, they  will couple to gravity.
If one solves the coupled Einstein-Higgs model, one again obtains a self-gravitating cosmic string, where the behaviour of
the $\Phi$ and $A_\mu$ is exactly the same as the coherence length and penetration length in the Ginzburg-Landau model.
The space time around the cosmic string far from the core, is Minkowski minus a wedge. The conical structure can be expressed
as an angle deficit $\Delta\varphi =8\pi G\mu$, where G is the gravitational constant and $\mu$ the mass density of
the string. The last decades many physicians  studied the consequences of topological defects in general relativity \cite{Vil}.
One of the earliest investigations was due to Marder \cite{Mar}, who described the gravitational Aharonov-Bohm effect in a conical space time.
The cosmic string can also be described as a point particle in (2+1) dimensional space time, because  the Killing vector $\frac{\partial}{\partial_z}$
is the axis of symmetry, and the angle deficit is expressed in the mass per unit length of the string.
An interesting example of the richness of the (2+1) dimensional gravity, is the Gott space time \cite{Gott}. An isolated pair of point
particles, moving with respect to each other, may generate a surrounding region where closed timelike curves (CTC) occur. It was not a surprise that one
can prove that the Gott space time has unphysical features \cite{Des1,Des2}.

It is quite natural to consider as a next step the non-Abelian Einstein-Yang-Mills situation is context with cosmic string solutions and
its unusual features like the formation of CTC's.
There is some evidence that, by suitable choice of the gauge, CTC's will not emerge dynamically \cite{Slag1,Slag2}.
Moreover, it is possible to write the model of the spinning EYM cosmic string as an effective U(1) Higgs model \cite{Slag1}.

In the most general setting, one also can add a cosmological constant ($\Lambda$) and a Gauss Bonnet term. Specially the influence of
a negative $\Lambda$ on the Gott space time was investigated \cite{Hol,Bir}. The influence of a GB term in a spherical symmetric 5-dimensional
EYM model was investigated recently \cite{Slag3}.

In this paper we investigate cosmic string features in a 5-dimensional space time. In section II we give an overview of the geometrical
properties of the (2+1) dimensional cosmon. In section III we outline our model. In section IV we present some numerical solutions of the model and
in section V we try to construct a Gott space time.

\section{Some history: the "cosmon"}

An interesting example of general relativity in  (2+1) dimensional space time, was the construction of a closed timelike curve (CTC)
by Gott \cite{Gott}. The space time generated by two moving cosmic strings (cosmons), or equivalently two point particles in (2+1) dimensions, would
produce a CTC when they move towards each other with sufficiently high relative velocity.

Several years later, Deser, Jackiw and 't Hooft found an elegant formulation of this so-called  Gott space time \cite{Des1,Des2}.
In fact in this Gott space time the CTC will also be present at spatial infinity, which constitutes an unphysical boundary condition.
Moreover, it turns out that the effective object has a tachyonic center of mass.
One can write the stationary spinning space time (orbital angular momentum) of the cosmic string (="cosmon") as
\begin{equation}
ds^2=-(dt+4 G{\cal J}d\varphi)^2+dz^2+dr^2+a^2 r^2d\varphi^2\quad (a>0),\label{eq1}
\end{equation}
with $a=1-4Gm=1-8\pi G\alpha$ and ${\cal J}$ constants. Here $\alpha$ represents the angle  deficit, ${\cal J}$ the angular momentum
and m the mass per unit length of the cosmic string. One can transform this metric by the transformation
\begin{equation}
T=t+4G{\cal J}\varphi,\qquad \varphi'=a\varphi,\label{eq2}
\end{equation}
into
\begin{equation}
ds^2=-dT^2+dz^2+dr^2+r^2d\varphi'^2.\label{eq3}
\end{equation}
This is Minkowski space time, with an angle deficit, because now $0\leq\varphi'\leq2\pi a$.
Moreover, this space time has a helical structure in time $T$: when $\varphi$ reaches $ 2\pi$, $T$ jumps by $8\pi G{\cal J}$, so we must
identify times which differ by $8\pi G{\cal J}$. The interval traced by a circle at constant $r$ and $t$
\begin{equation}
\Delta s^2=(a^2r^2-16 G^2{\cal J}^2)<0,\label{eq4}
\end{equation}
is timelike for $r<\frac{4G{\cal J}}{a}$. The question is whether this will happen in our universe. Can the cosmon be confined within a small enough region
to satisfy ${\cal J}>\frac{ar_0}{4G}$? It will never occur in a finite time. To prove the conjectures above, one considers the cosmon in the
(2+1)-dimensional space time by dropping the $dz^2$ term.
To close the space around the "particle", one has the matching conditions for identifying points $(\vec x',\vec x)$ along the
edges by $\vec x'=\Omega(\beta)\vec x $, with $\Omega$ and  the rotation matrix in 2 dimensions and  $\beta =4\pi Gm =\pi(1-a)$. For two particles
at the origin and  at $\vec a$ the conditions consist of two rotations to close the space:
$\vec x'=\Omega (\beta_1)\vec x, \vec x''=\Omega_1(\beta_1)(\vec a +\Omega_2(\beta_2)(\vec x -\vec a))=\vec b +\Omega_1\Omega_2(\vec x -\vec b)$.
So one has effectively one particle at $\vec b$.
For the one particle situation, we had the condition $a>0$, or $m<\frac{1}{4G}$. If $a<0$ the metric around the particle becomes singular,i.e.,
the distance from the particle to any other point diverges as $r^a$. This can be seen by considering the general axial symmetric 2-space
\begin{equation}
ds^2=e^{2N}d\rho^2+\rho^2d\varphi^2.\label{eq5}
\end{equation}
Transforming to our conical form yields
$\rho=e^{-N}r,\varphi=e^{-N}\varphi',r=\frac{R^a}{a}, \varphi'=a\varphi''$.
So $a<0$ is physically hard to accept.
In the Gott space we have two moving particles located at $\vec a_1$ and $\vec a_2$.
 The matching condition consists now of  two rotations $\Omega_1, \Omega_2$
and  two boosts $L_1, L_2$. The effective particle
( with a center of mass) can be presented as $\vec x''=\Omega_{eff}\vec x +\vec c$ with $\Omega_{eff}=L_1\Omega_1L_1^{-1} L_2\Omega_2 L_2^{-1}$ and
$\vec c=L_1\Omega_1L_1^{-1}(\vec a_2-\vec a_1)$
We are at the center of mass if $\Omega_{eff}$ is purely spacelike, i.e., of the form ${\cal L}\Omega_M {\cal L}^{-1}$ with $\Omega_M$ a pure rotation.
and with $\vec c$ timelike. It turns out \cite{Des2} that the time component of $\vec c$ is just the angular momentum. A possible spacelike component
of $\vec c$ would imply that the effective spinning particle is not at the origin.
Taking the trace of the transformations we obtain:
\begin{equation}
{\bf Tr} \Omega_M=1+2\cos 8\pi GM={\bf Tr}(L_1\Omega_1L_1^{-1}L_2\Omega_2L_2^{-1}).\label{eq6}
\end{equation}
From $\cos8\pi GM <1$ we then obtain
\begin{equation}
\cosh\xi \sin 4\pi Gm<1,\label{eq7}
\end{equation}
with $\xi$ the rapidity, such that $\cosh\xi=\frac{1}{\sqrt{1-v^2}}$ and $\tanh\xi=v$ ($v$=velocity).
Gott's construction demands the opposite, so M is imaginary and the identification is boostlike. The conical structure
of the $(x,y)$-plane is replaced by one in the $(t,\vec v)$-plane plus a jump in the remaining spatial direction.
In general: A boost-identified space time will never arise by boosting a rotation-identified space time, so
the effective object has a tachyonic center of mass. One can remove the line-like obstruction, but then the space time
becomes periodic in time and CTC's are present. If one wants to avoid this, one must keep the obstruction and its precise
location is arbitrary. So the Gott-pair is surrounded by a boundary condition that CTC's are also at infinity.

The  prove was not complete, because one could wonder what will happen in a closed spacetime. It was proposed that Gott's
condition can be fulfilled, if a heavy particle decays into two lighter ones \cite{Car}. However, one can also prove \cite{Hooft}
that in this model a causal situation is present. If one follows the particles in a time-ordered manner, then the lifetime of
the system is finite and the 2-volume of this universe decreases with time until a big crunch ends it all.

\section{The generalized model in 5-dimensional space time}
Let us now consider a 5-dimensional space time
The action of the model under consideration is
\begin{equation}
{\cal S}=\frac{1}{16\pi}\int d^5x\sqrt{-g_5}\Bigl[\frac{1}{ G_5}(R-\Lambda)-\frac{1}{g^2}Tr{\bf F^2}\Bigr],\label{eq8}
\end{equation}
with $G_5$ the gravitational constant, $\Lambda$ the cosmological constant and $g$ the gauge coupling.
The coupled set of equations  will then become
\begin{eqnarray}
\Lambda  g_{\mu\nu}+G_{\mu\nu} =8\pi G_5 T_{\mu\nu}, \label{eq9}
\end{eqnarray}
\begin{eqnarray}
{\cal D}_\mu F^{\mu\nu a}=0,\label{eq10}
\end{eqnarray}
with the Einstein tensor
\begin{eqnarray}
G_{\mu\nu}= R_{\mu\nu}-\frac{1}{2}g_{\mu\nu} R. \label{eq11}
\end{eqnarray}
Further, with
$R_{\mu\nu}$ the Ricci tensor and $T_{\mu\nu}$ the energy-momentum tensor
\begin{eqnarray}
T_{\mu\nu}={\bf Tr}F_{\mu\lambda}F_\nu^\lambda -\frac{1}{2}g_{\mu\nu}{\bf Tr}F_{\alpha\beta}F^{\alpha\beta},\label{eq12}
\end{eqnarray}
and with $F_{\mu\nu}^a=\partial_\mu A_\nu^a -\partial_\nu A_\mu^a +g\epsilon^{abc}A_\mu^b A_\nu^c $, and
${\cal D}_\alpha F_{\mu\nu}^a=\nabla_\alpha F_{\mu\nu}^a+g\epsilon^{abc}A_\alpha^b F_{\mu\nu}^c$
where $A_\mu^a$ represents  the YM potential.

Consider now the stationary axially symmetric  5-dimensional space time
\begin{equation}
ds^2=-F(r)(dt+\omega(r)d\psi)^2+dr^2+dz^2+A(r)^2r^2d\varphi^2+B(r)^2d\psi^2, \label{eq13}
\end{equation}
with the YM parameterization only in the brane:
\begin{eqnarray}
A_t^{(a)}=\Bigl(\Phi(r)\cos\varphi,\Phi(r)\sin\varphi,0\Bigr), A_r^{(a)}=A_z^{(a)}=A_\psi^{(a)}=0,\cr
A_\varphi^{(a)}=\Bigl(0,0 ,W(r)-1\Bigr).\label{eq14}
\end{eqnarray}
To derive a set of differential equations from the field equations Eq.(\ref{eq9}) and Eq.(\ref{eq10}), we used Maple.
First of all, one can take the combination of the $(t,t)$ and $(\psi ,t)$ components of the Einstein equations, which yields
the equation for $\omega''$. From the combination of the $(t,t)$ and $(z,z)$ components we have an equation for $F''$ and the equations
for $A''$ and $B''$ follow from the combination of the $(t,t), (z,z)$ and $(\varphi,\varphi)$ components.
From the $(2,\psi)$ and $(3,\varphi)$ components of the Yang-Mills equations Eq.(\ref{eq10}) we obtain equations for $W''$ and $\Phi''$.
The results are in this way (we take $g=1$)
\begin{eqnarray}
A''=-A'\Bigl(\frac{B'}{B}+\frac{2}{r}+\frac{F'}{2F}\Bigr)-\frac{A}{r}\Bigr(\frac{B'}{B}+\frac{F'}{2F}\Bigr)\cr
-\frac{8\pi G}{r^2FAB^2}\Bigr(\Phi^2W^2(\omega^2F-B^2)+FB^2(W')^2\Bigr),\label{eq15}
\end{eqnarray}
\begin{eqnarray}
B''=\frac{BF'}{2F}\Bigl(\frac{1}{r}+\frac{A'}{A}\Bigr)+\frac{3F(\omega ')^2}{4B} -\frac{1}{2}\Lambda B \cr +\frac{4\pi G}{r^2FA^2B}\Bigl(
FB^2(W')^2+\Phi^2W^2(\omega^2F-3B^2)-A^2r^2(\Phi ')^2(\omega^2F+B^2)\Bigr),\label{eq16}
\end{eqnarray}
\begin{equation}
F''=F'\Bigl(\frac{F'}{2F}-\frac{B'}{B}-\frac{A'}{A}-\frac{1}{r}\Bigr)-\frac{F^2(\omega')^2}{B^2}
+\frac{16\pi G}{r^2A^2}\Bigl(A^2r^2(\Phi')^2+\Phi^2W^2\Bigr),\label{eq17}
\end{equation}
\begin{equation}
\omega''=\omega'\Bigl(\frac{B'}{B}-\frac{A'}{A}-\frac{1}{r}-\frac{3F'}{2F}\Bigr)-\frac{16\pi G\omega}{r^2A^2F}\Bigl(
A^2r^2(\Phi')^2+\Phi^2W^2\Bigr),\label{eq18}
\end{equation}
\begin{equation}
\Phi''=\Phi'\Bigl(\frac{B'}{B}-\frac{A'}{A}-\frac{F'}{2F}-\frac{1}{r}-\frac{\omega'}{\omega}\Bigr)+\frac{\Phi W^2}{A^2 r^2},\label{eq19}
\end{equation}
and
\begin{equation}
W''=W'\Bigl(\frac{1}{r}-\frac{B'}{B}+\frac{A'}{A}-\frac{F'}{2F}\Bigr)+\frac{\Phi^2W}{B^2F}\Bigl(\omega^2F-B^2\Bigr).\label{eq20}
\end{equation}

From a combination of the $(1,t)$ and $(1,\psi)$ components of the  YM equations, we obtain also for the angular momentum component $\omega$
a first order  expression
\begin{equation}
\omega'=\frac{B\omega (2FB'-BF')}{F(F\omega^2+B^2)},\label{eq21}
\end{equation}
which can be used in the numerical code.

We also have a constraint equation. From the $(\psi,\psi)$ component of the Einstein equations, we obtain
\begin{equation}
\Lambda FB^2A^2r^2=8\pi G\Bigl[(\omega^2F-B^2)\Bigl(\Phi^2W^2+A^2r^2(\Phi')^2\Bigr)+FB^2(W')^2\Bigr].\label{eq22}
\end{equation}
From this equation  we obtain for the $g_{\psi\psi}$ component
\begin{equation}
g_{\psi\psi}=\frac{FB^2(W')^2-\frac{\Lambda FB^2A^2r^2}{8\pi G}}{(\Phi^2W^2+A^2r^2(\Phi')^2)}.\label{eq23}
\end{equation}
So a negative cosmological constant will keep $g_{\psi\psi}$ positive, which is desirable.
Further, from a combination of $\nabla_\mu T^{\mu\nu}=0,$ and the $(3,\varphi), (2,\psi)$ components of the YM equations,
we obtain the same equation as Eq.(\ref{eq21}) which proves the consistency of the system.

\section{Numerical solutions}
\subsection{Initial value solutions}
For any given set of initial conditions, the differential equations determine the behaviour of the YM- and metric fields.
In the Abelian U(1) cosmic string model, one is particular interested in solutions with asymptotic conditions for the gauge field and scalar field,
i.e., the gauge field approaching zero and the scalar field approaching to 1 far from the string.
In our non-Abelian rotating case, the asymptotic behaviour is far from clear. For sure, one would like to see an
asymptotic flat space time minus a wedge. So one can try to find "acceptable solutions" by shooting and fine tuning the
initial conditions\cite{Dy,Lag}.

We will take for the initial values of the YM gauge fields $W$ and $\Phi$ the usual ones, i.e., $W(0)=1, \Phi(0)=0$
and choose for $F(0)=1$.
So we have a set of  initial parameters and 2 fundamental constants $\Lambda, G_5$.

The equations are easily solved with an ODE solver and checked  with an ODE solver in MAPLE.

In figure 1 we plotted the metric components, the YM components, $g_{\psi\psi}$ and the Kretschmann curvature invariant
 for two different initial values of $\omega, \omega'$ and for $\Lambda =0$.
We observe that when $g_{\psi\psi}$ approaches zero and even becomes negative, the solution becomes singular.
In fact the Kretschmann curvature-invariant blows up
to minus infinity, as expected. So it is not likely that CTC's will form.
In the angle variable $\varphi$ there is an angle deficit
\begin{equation}
\Delta\varphi =2\pi\Bigl(1-\lim_{r\to\infty}\partial_r\sqrt{g_{\varphi\varphi}}\Bigr)\label{eq24}
\end{equation}
present. In figure 1 $\Delta\varphi$ approaches a value $<2\pi$ (before the singularity), so the space is conical outside the string.
If we include a negative $\Lambda$, we obtain for example figure 2. We observe that the oscillatory behaviour of $W$ resolves and
$g_{\psi\psi}$ never approaches zero. The Kretschmann scalar approaches to zero. Further, we see that $\omega \rightarrow 0$  far
from the string's core, as desired. There is still an angle deficit. However is is hard to call this solution a 
global Nielsen-Olesen Abelian type of cosmic string because $\Phi$ is unbounded for large r. 

\begin{figure*}
\centerline{
 \hfill \includegraphics[width=0.35\textwidth]{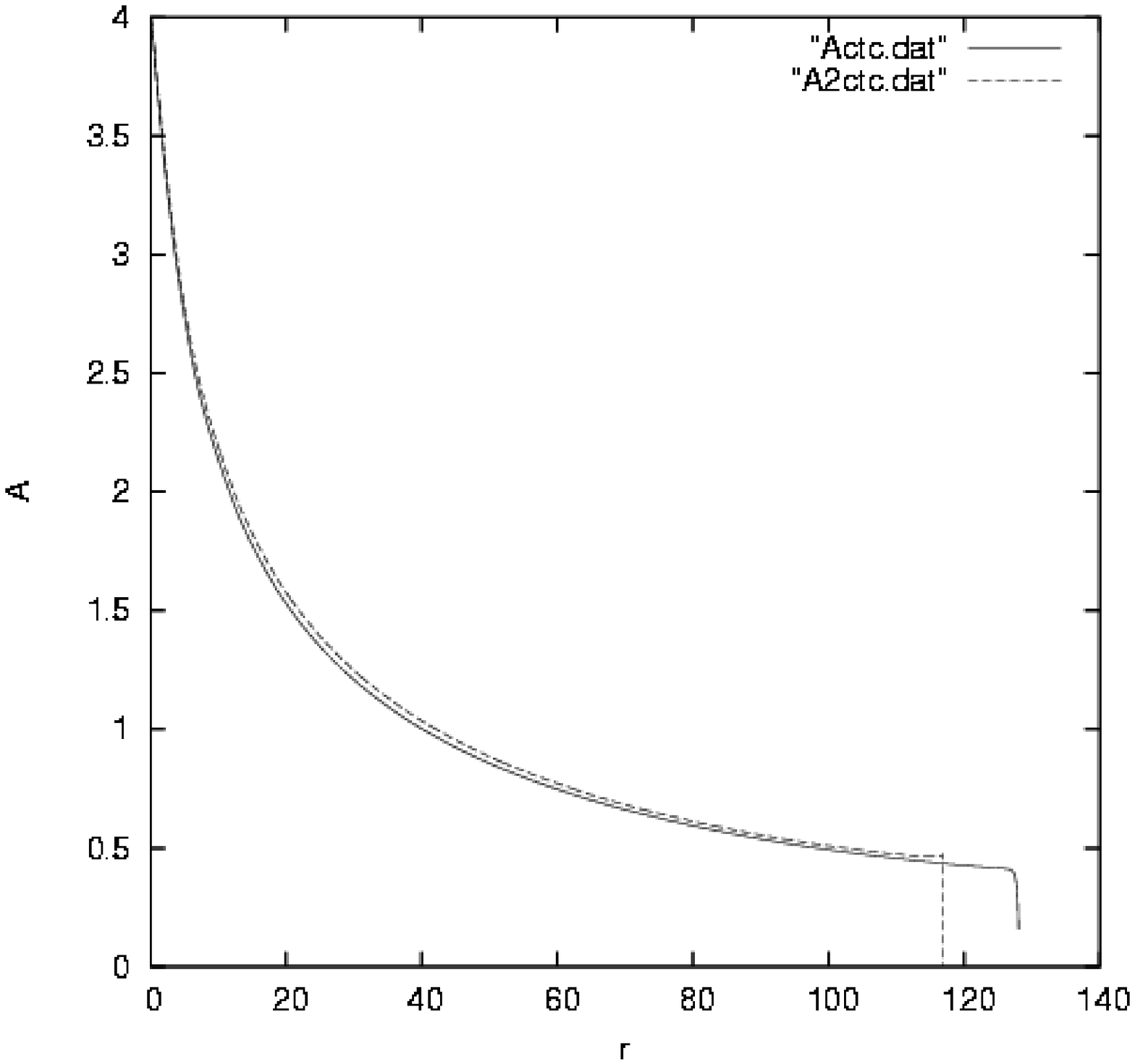}
 \includegraphics[width=0.35\textwidth]{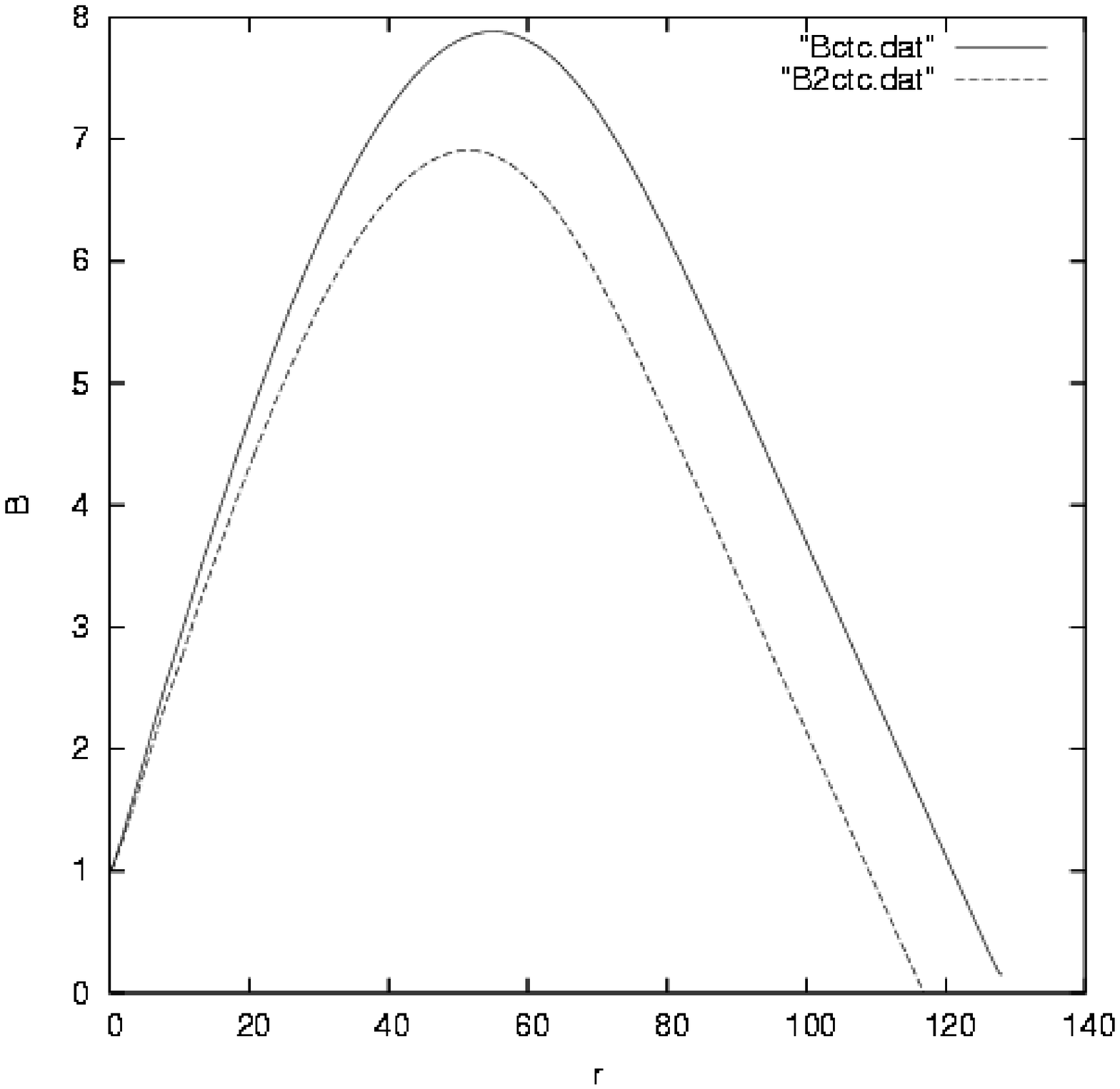}
 \includegraphics[width=0.35\textwidth]{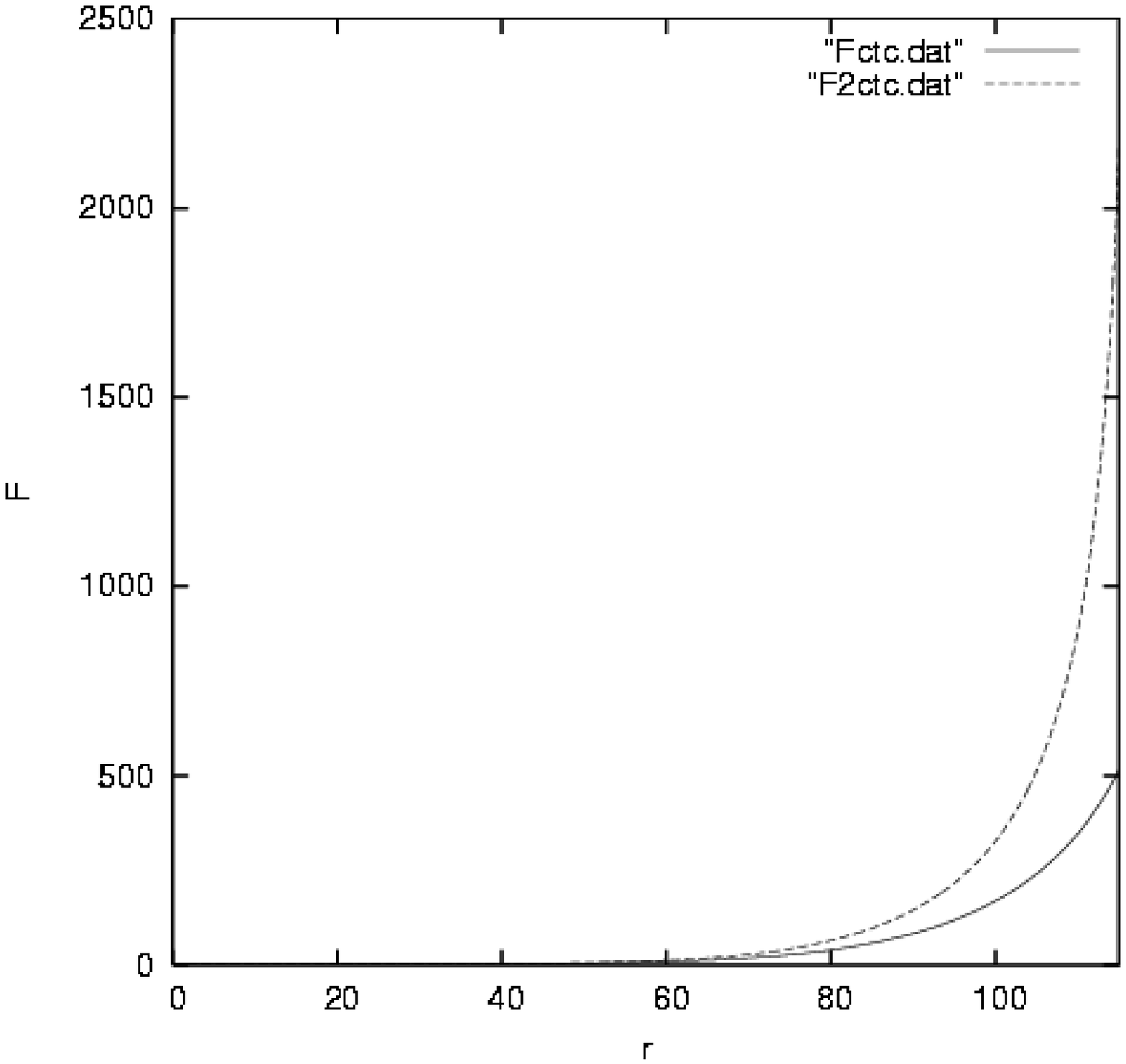}
\hfill}
\centerline{
\hfill
 \includegraphics[width=0.35\textwidth]{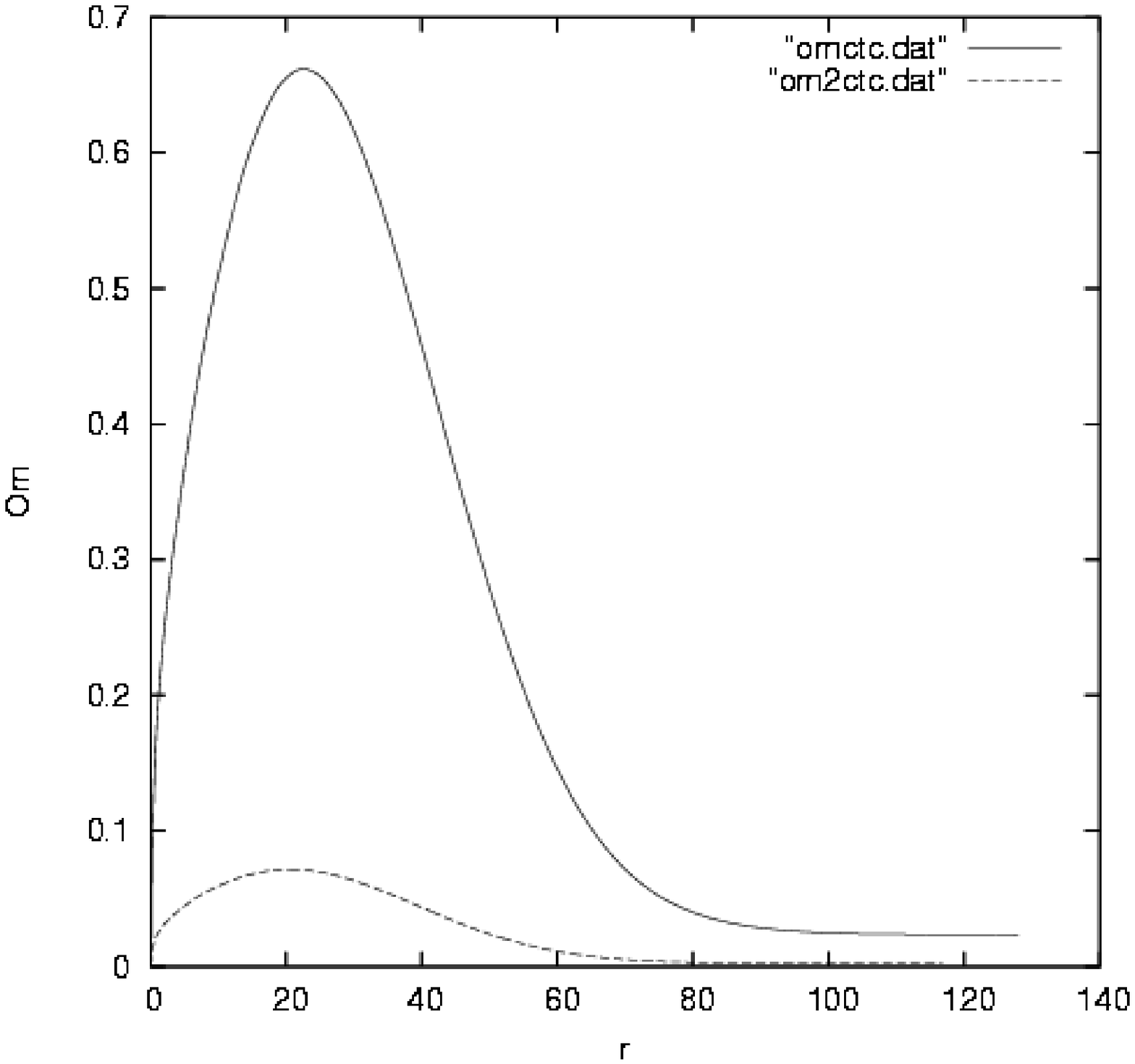}
\includegraphics[width=0.35\textwidth]{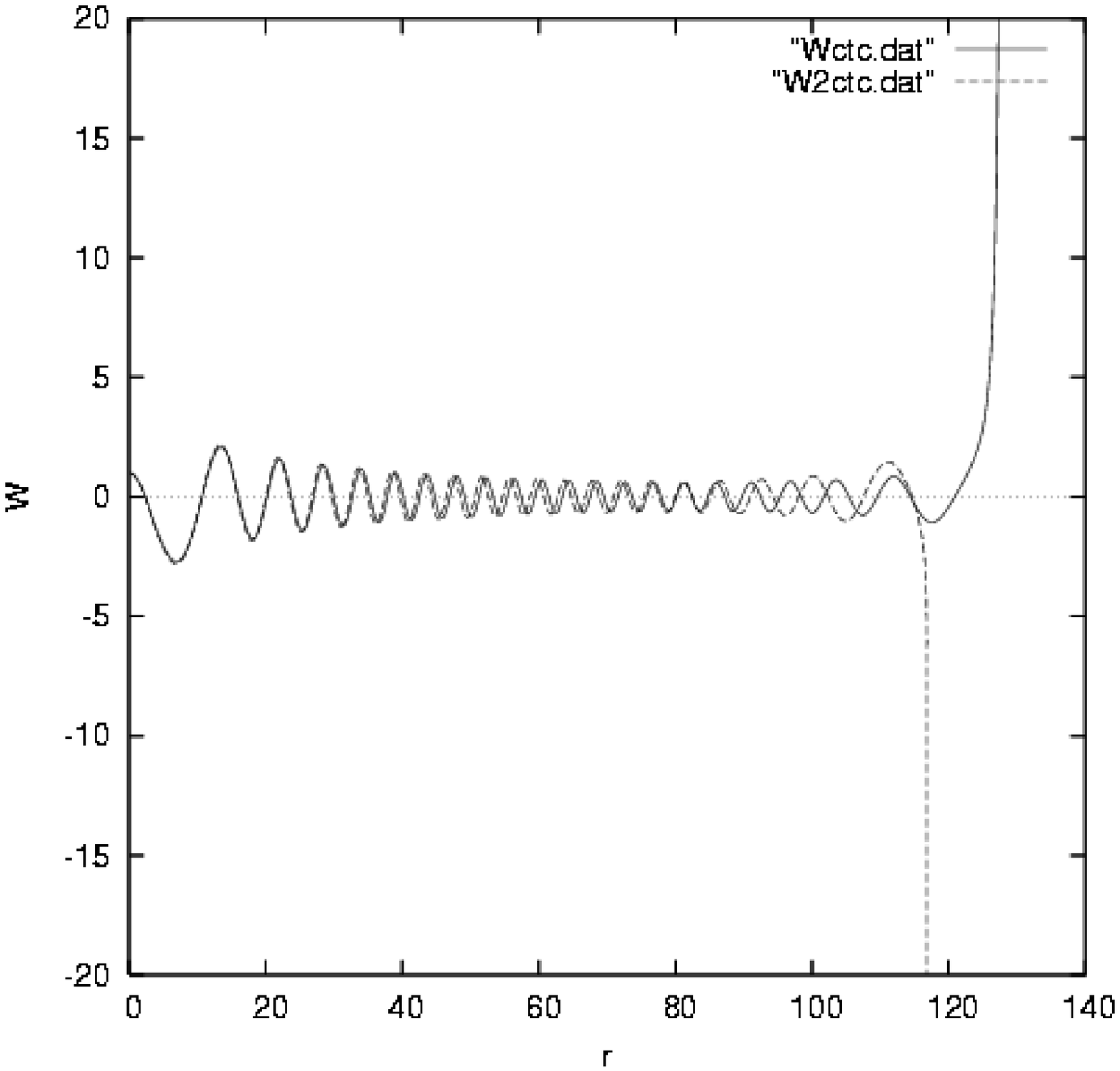}
\includegraphics[width=0.35\textwidth]{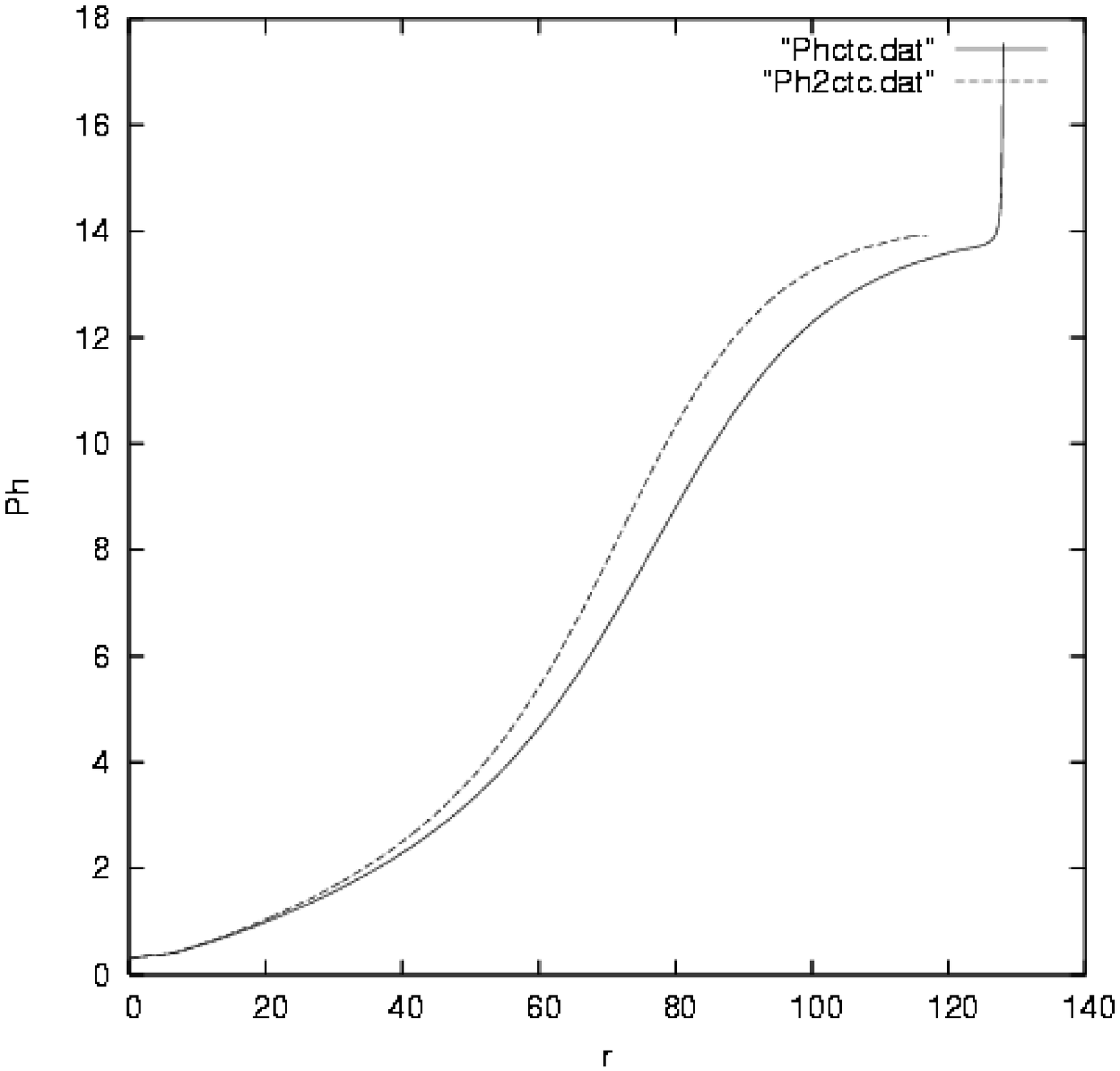}
\hfill}
\centerline{
\hfill \includegraphics[width=0.35\textwidth]{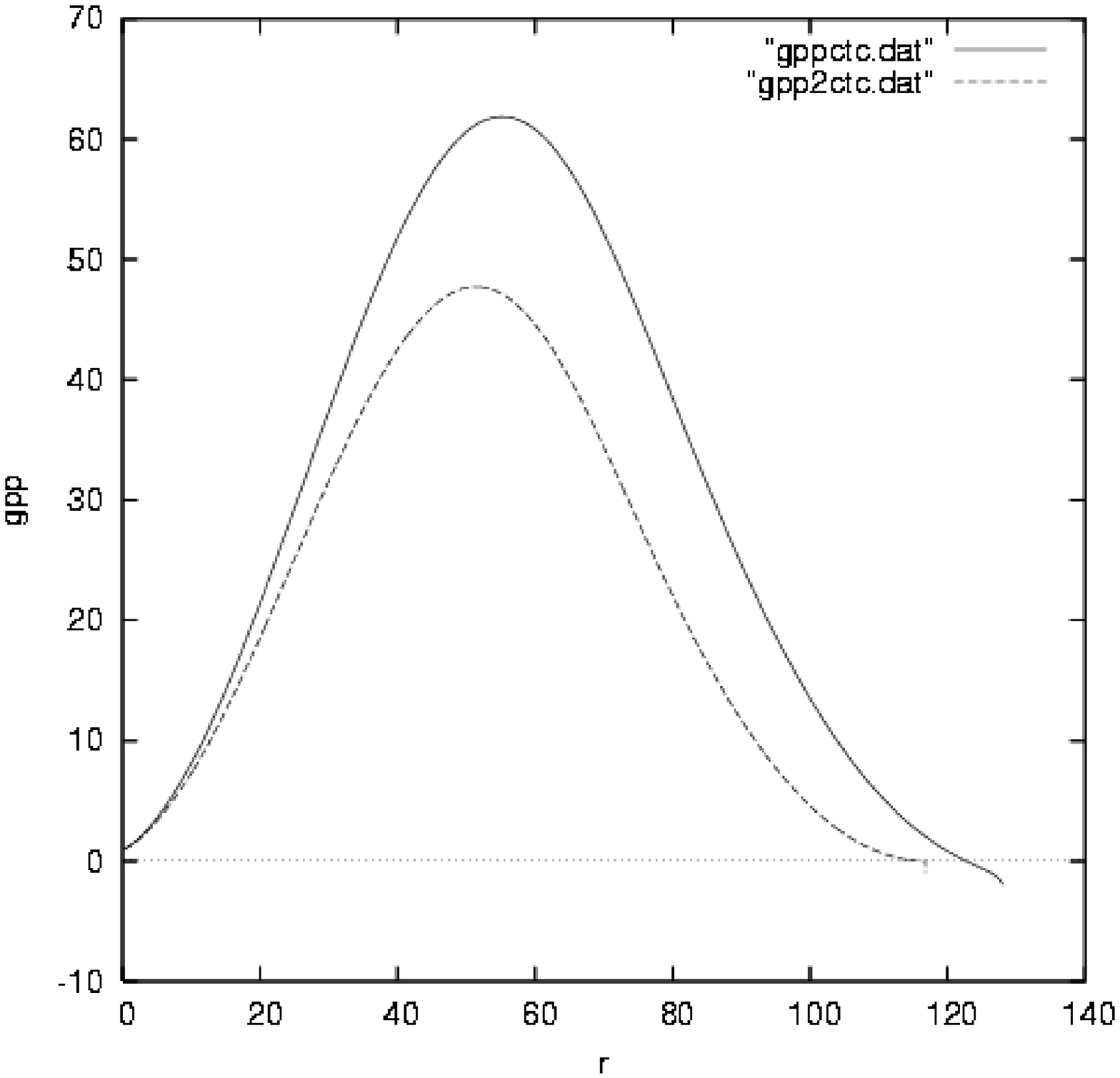}
\includegraphics[width=0.35\textwidth]{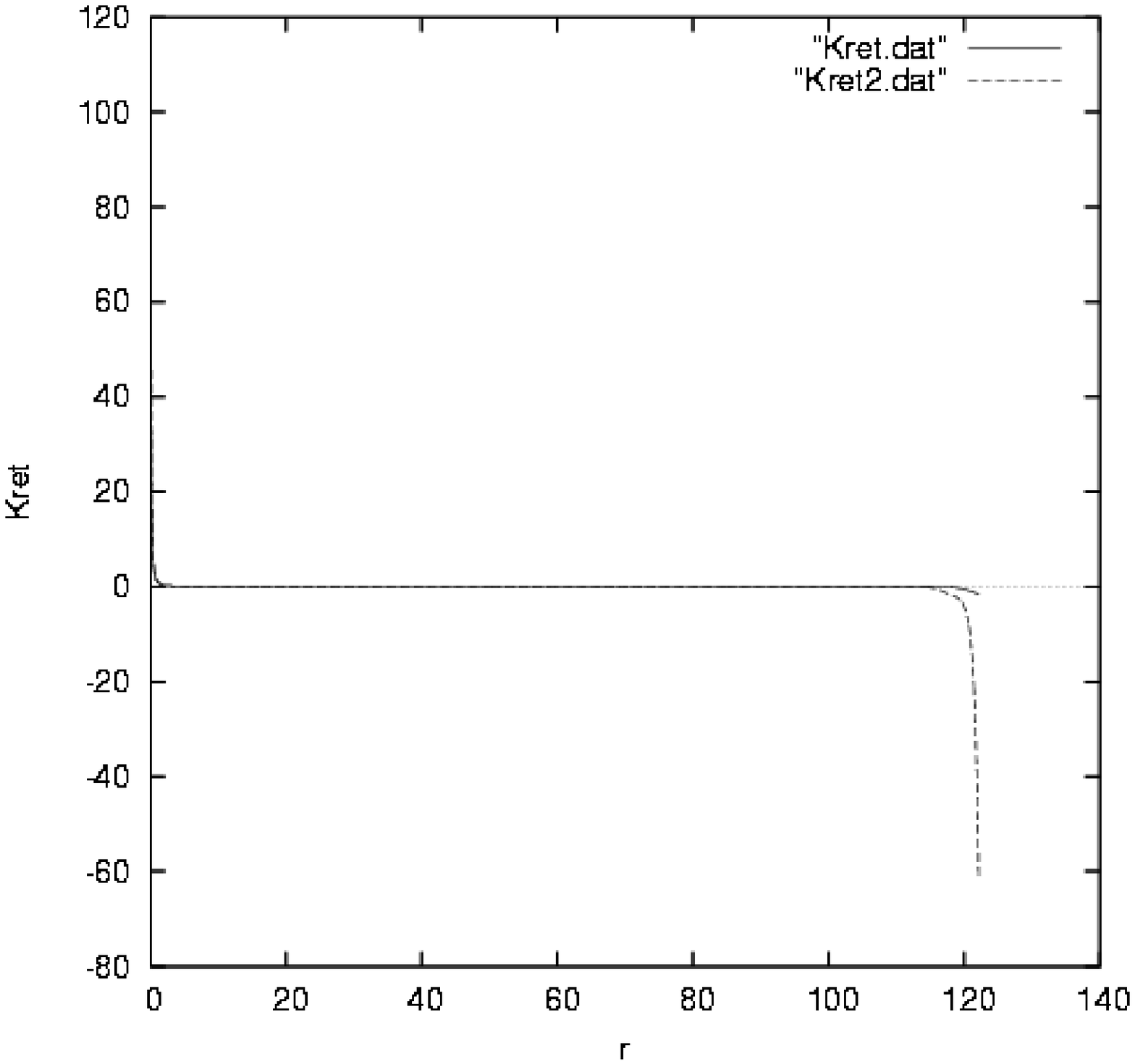}
\hfill}
\caption{ Two typical solutions for the metric- and YM-components for two slightly different initial values for $\omega(0)$ and $\omega'(0)$.
Solid line: $\omega(0)=0.0001, \omega'(0)=0.8$; dotted line: $\omega(0)=0.001, \omega'(0)=0.1$. Further: A(0)=4, B(0)=1, F(0)=1, W(0)=1,
    $\Phi(0) =0.3$, A'(0)=0, B'(0)=0, F'(0)=0.4, W'(0)=-0.03, $\Phi'(0)=0$, G=0.5, $\Lambda=0$. We also plotted $g_{\psi\psi}$ and
    the Kretschmann scalar. The Kretschmann scalar  diverges when $g_{\psi\psi}$ approaches zero. There is also an angle deficit.}
\label{fig:1}
\end{figure*}

\begin{figure*}
\centerline{
 \hfill \includegraphics[width=0.35\textwidth]{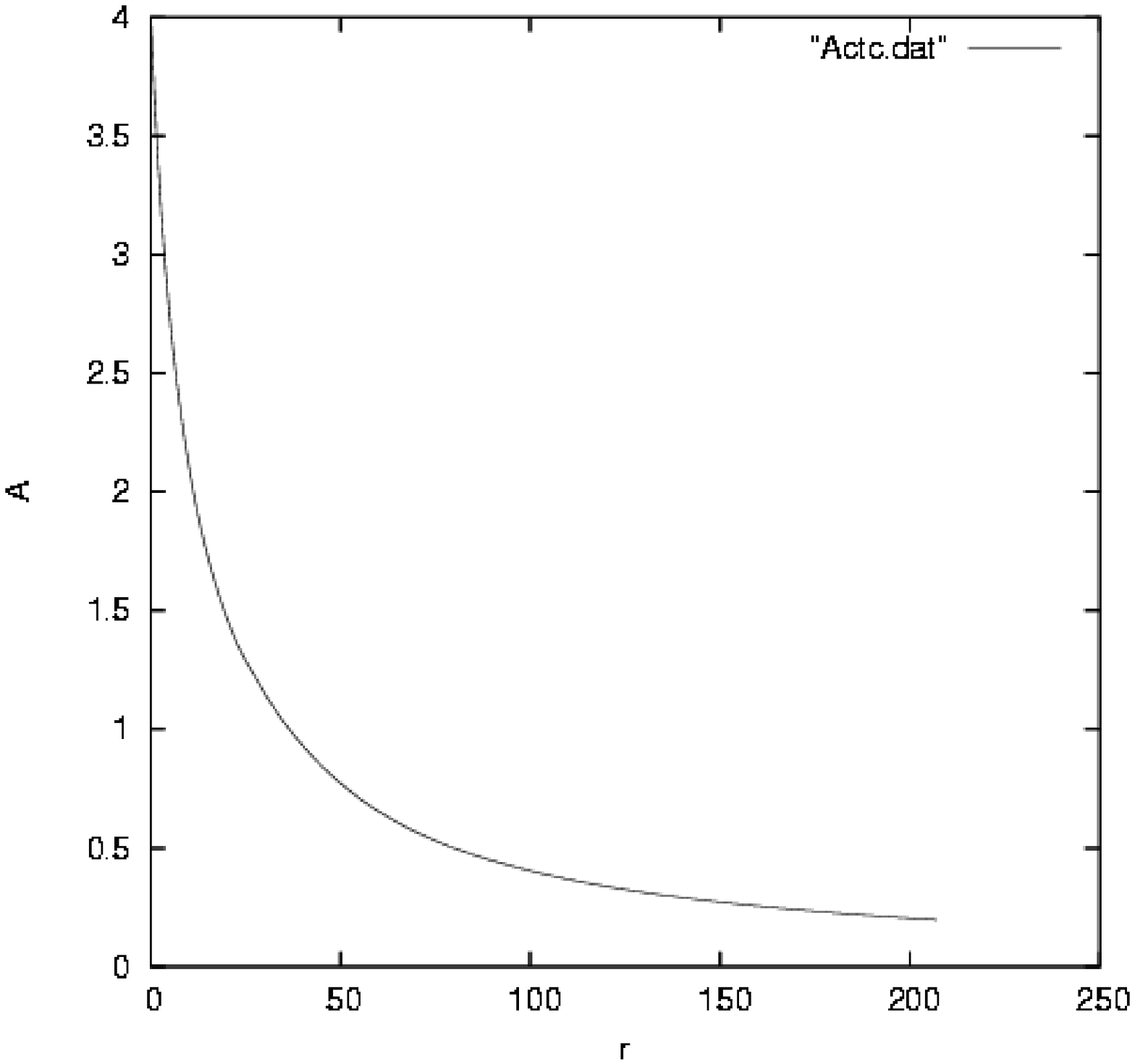}
 \includegraphics[width=0.35\textwidth]{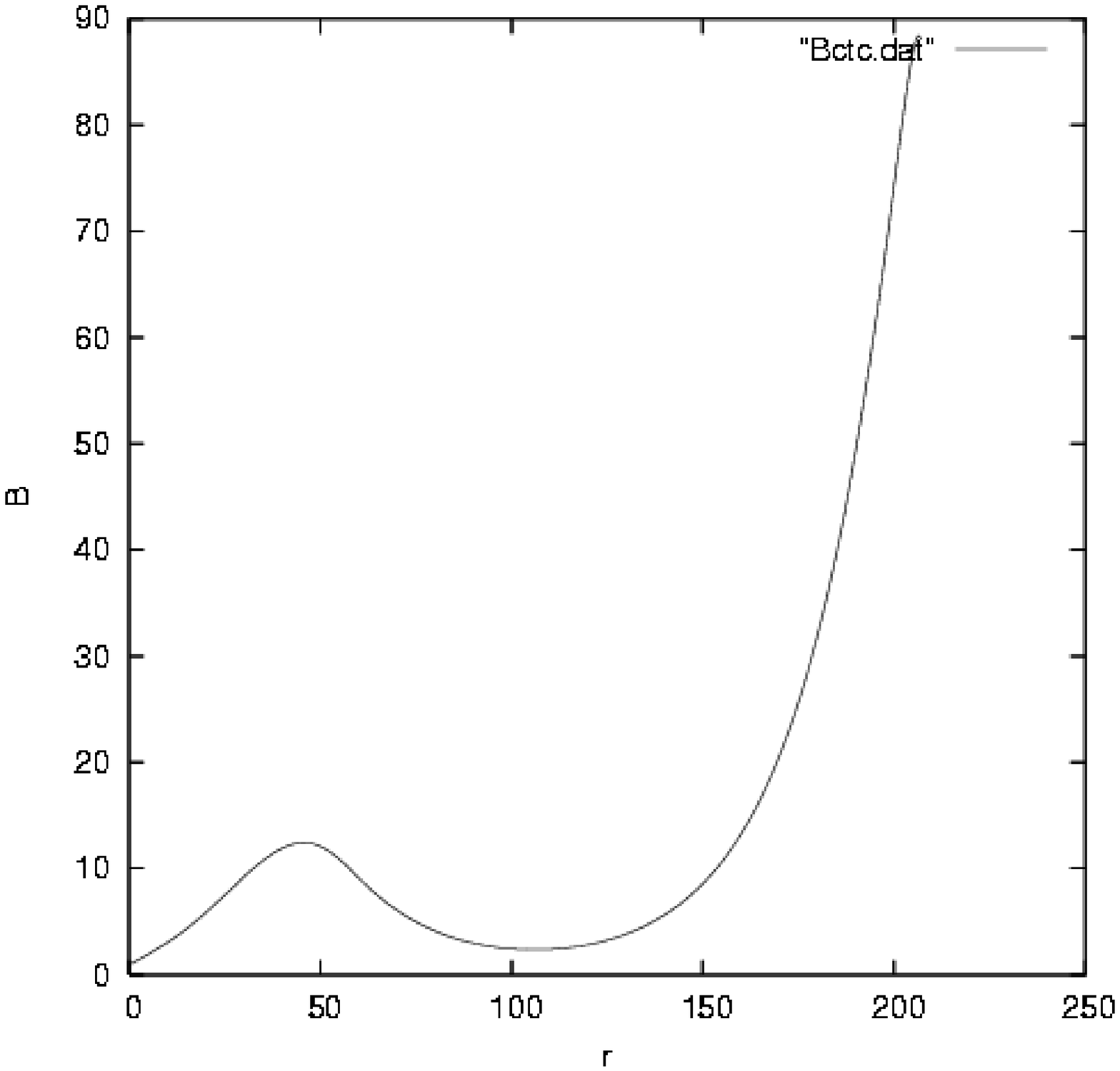}
 \includegraphics[width=0.35\textwidth]{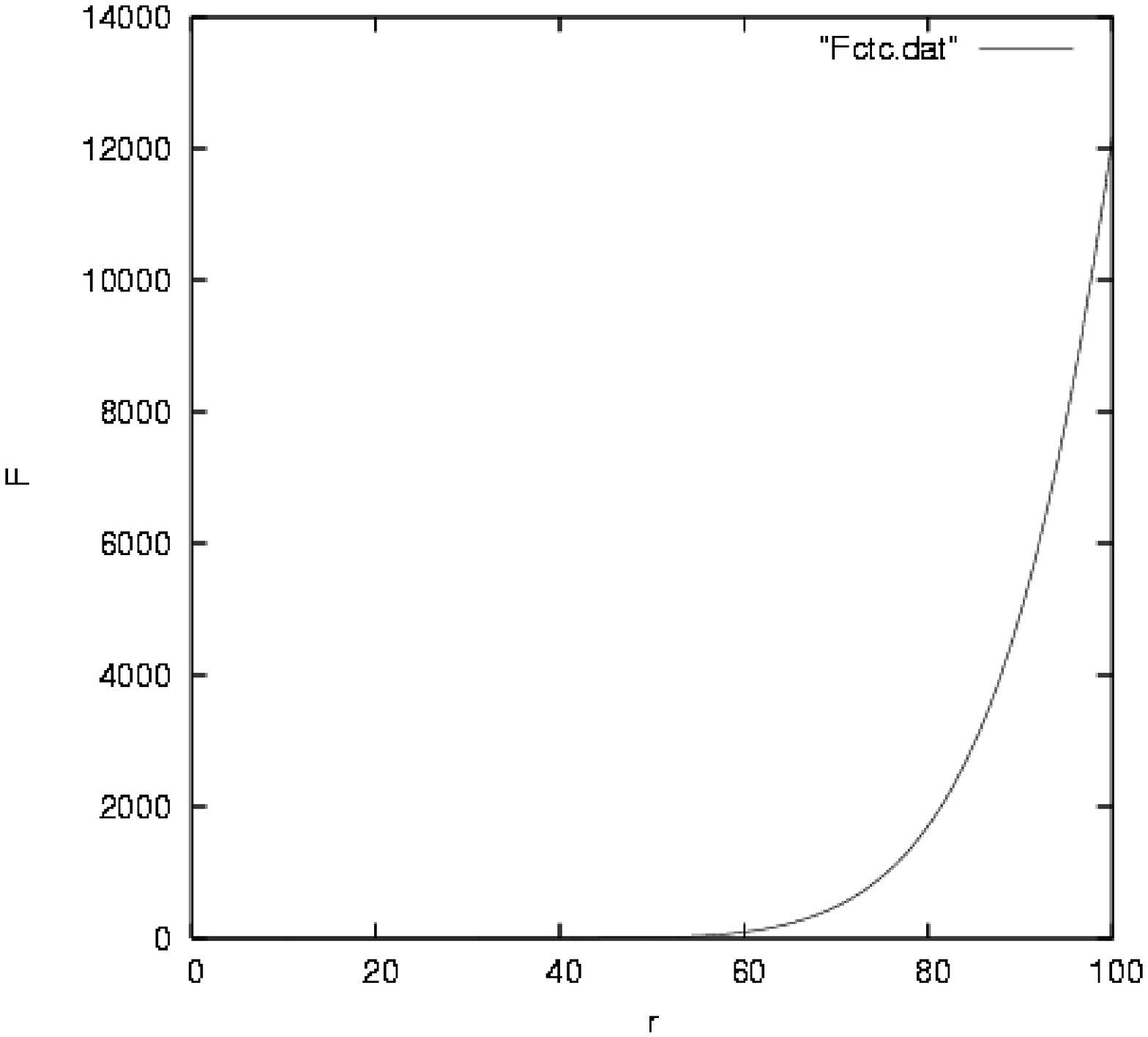}
\hfill}
\centerline{
\hfill
 \includegraphics[width=0.35\textwidth]{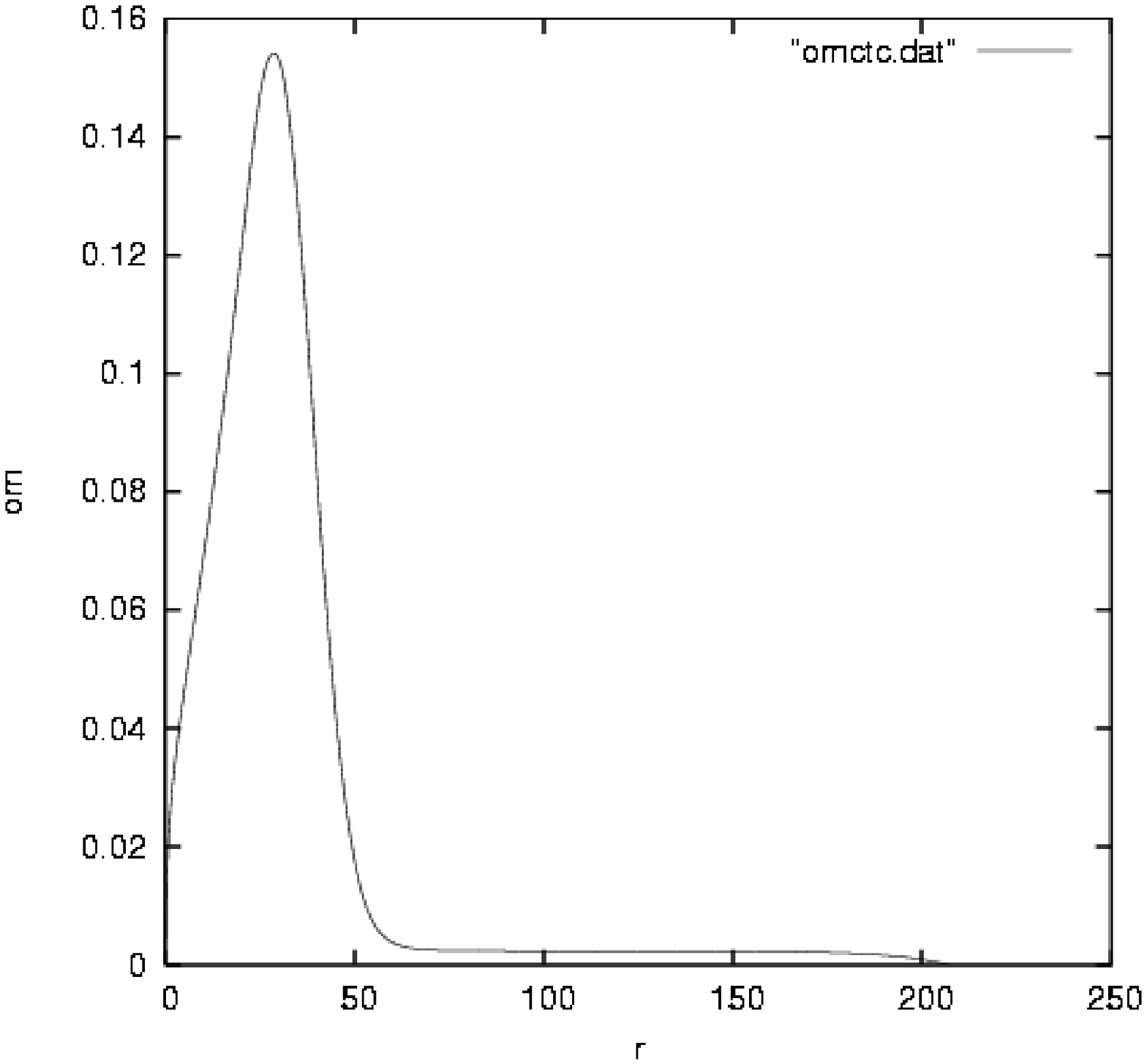}
\includegraphics[width=0.35\textwidth]{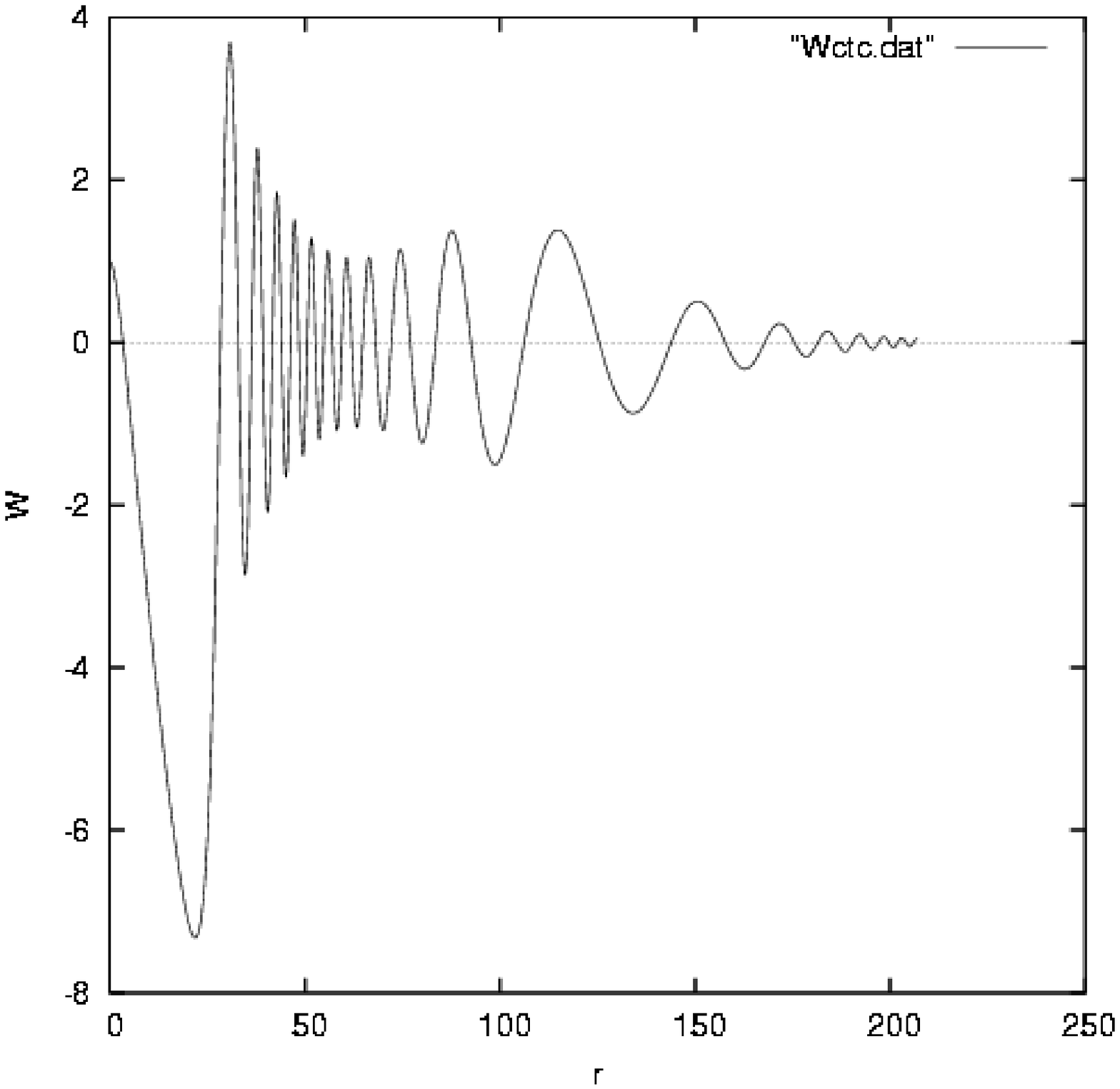}
\includegraphics[width=0.35\textwidth]{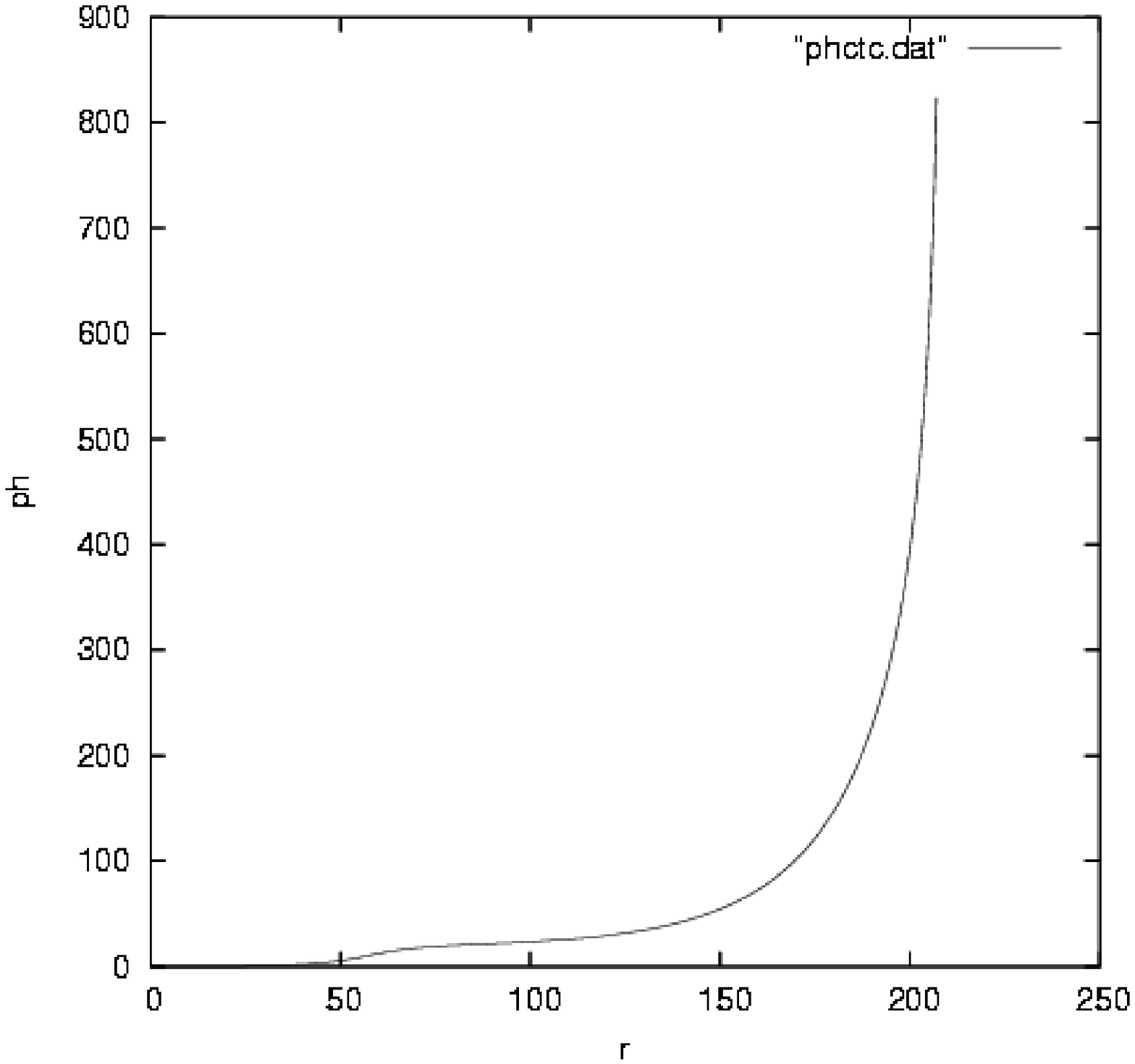}
\hfill}
\centerline{
\hfill \includegraphics[width=0.35\textwidth]{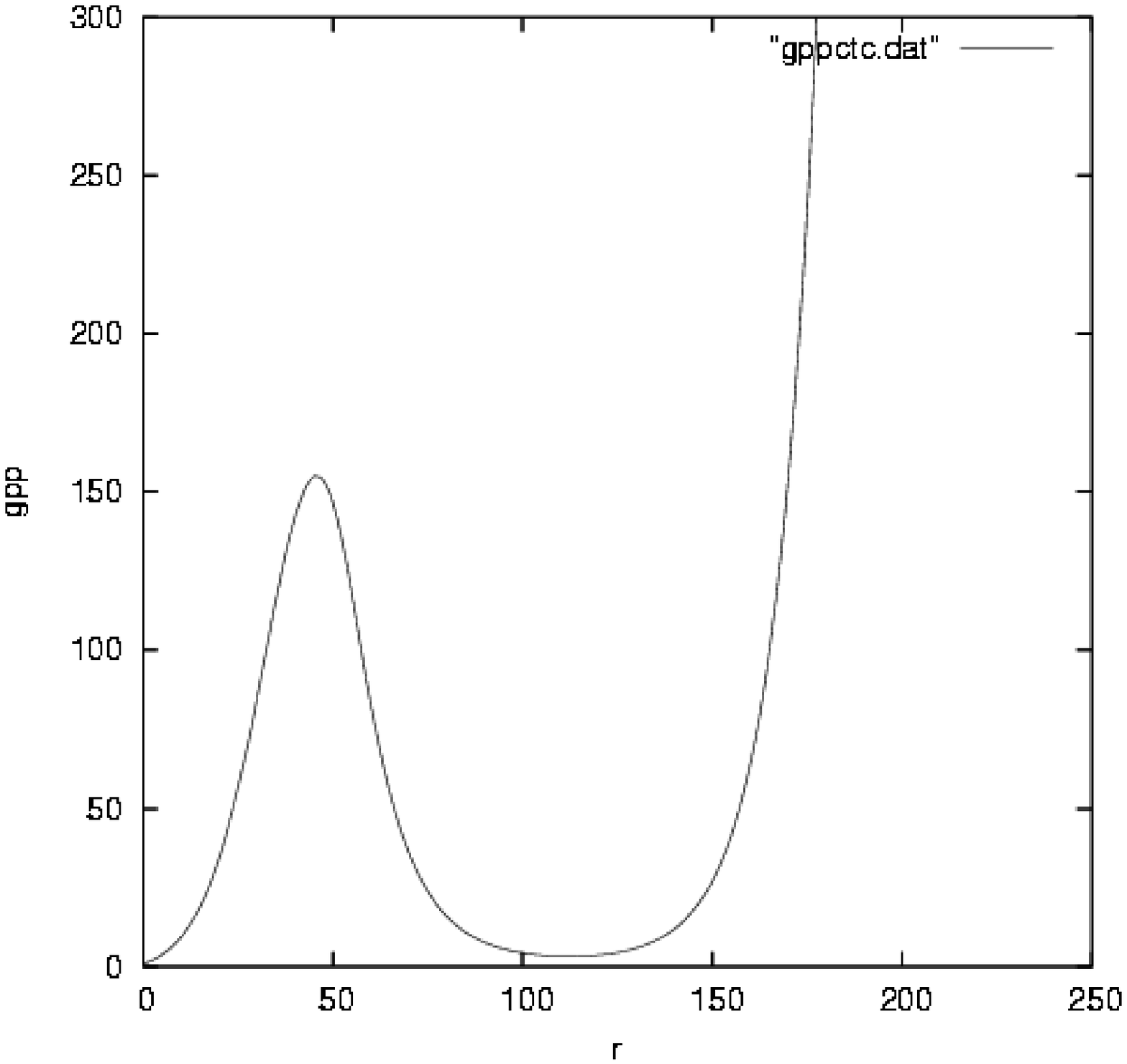}
\hfill}
\caption{Typical solution for negative $\Lambda$. Initial values: A(0)=4, B(0)=1, F(0)=1, $\omega(0) =0.0001$,W(0)=1, $\Phi(0)=0.015$
         A'(0)=0, B'(0)=0, F'(0)=0.4, $\omega'(0)=0.1$, W'(0)=-0.03, $\Phi'(0)=0.03$, G=0.5, $\Lambda=-0.004$. The Kretschmann scalar now
         becomes zero for large r. The solution is regular everywhere with an angle deficit.}
\label{fig:2}
\end{figure*}

\subsection{Two point boundary value solutions}
In order to fulfil the asymptotic behaviour of $W$ and $\Phi$, we use a different numerical method.

In figure 3 we plotted a two point boundary value solution for a positive and negative value of $\Lambda$ and for boundary values 
for the two YM components $W$ and $\Phi$. 
We took the boundary conditions of a Nielsen-Olesen string: $W=0$ and $\Phi=1$ at the endpoint of r and let 
the program vary the initial values at r=0. It is known that in the Abelian counterpart model\cite{Laguna} the gauge fields behave 
asymptotically like
\begin{equation}
\Phi=1-c_1e^{-r},\quad w=c_2e^{-c_3r},\label{eq25}
\end{equation}
with $c_i$ some constants.So $r\approx 20$ for  the right boundary is sufficient large. 

For positive  $\Lambda$ we observe that W tends to grow to a large negative value. So when we insist on the Nielsen-Olesen boundary conditions for 
large $r$, negative $\Lambda$ is more likely in order to keep W between $\pm 1$. The oscillatory behaviour of $W$ is not uncommon 
for gravitating YM vortices\cite{Vol2} and  is due to the moving of the YM particle in the inverted double well potential.
It is obvious that again
a negative value of $\Lambda$ is more likely to obtain a regular behaviour of $W$ with correct string like behaviour for large r.

In figures 4 and 5 we plotted  2 solutions with right boundary conditions $\Phi'=W'=0$, for $\Lambda=0$. The initial condition for $A$ differs
slightly. We see in figure 5 a non-regular solution, i.e., $W$ grows unlimited.
In figure 6 we plotted a regular solution for negative $\Lambda$. $W$ is oscillating around zero with decreasing amplitude and
there is a angle deficit.
In figure 7 we plotted a solution for positive $\Lambda$. Now a CTC has formed ($g_{\psi\psi}<0$).

So again a negative $\Lambda$ is favourable for regularity.

Similar result was found from stability analysis in the particle-like solutions in the spherically symmetric  
5-dimensional EYM model\cite{Okuy}: stable solutions only exist for negative $\Lambda$. 
However, in this model, only a magnetic component is present, which results in a 
so-called "quasi-local" asymptotically AdS space time because the ADM mass is not finite.

We can try to find an asymptotic form of the metric component $A$, when we choose for the YM fields the asymptotic  forms of the Eq.(\ref{eq25}). 
When $\omega=0$ and  $F=c_4+c_5r^2$ for large r ($c_i$ some constants), then from the field equations it follows that $A(r)$ approaches 
a constant value for large r and for some values of the constants $c_i$. So there is a conical structure outside the core of the  string.
This behaviour in higher dimensional gravity models with  non-Abelian gauge fields is not uncommon\cite{Nak}.

\begin{figure*}
\centerline{
 \hfill \includegraphics[width=0.55\textwidth]{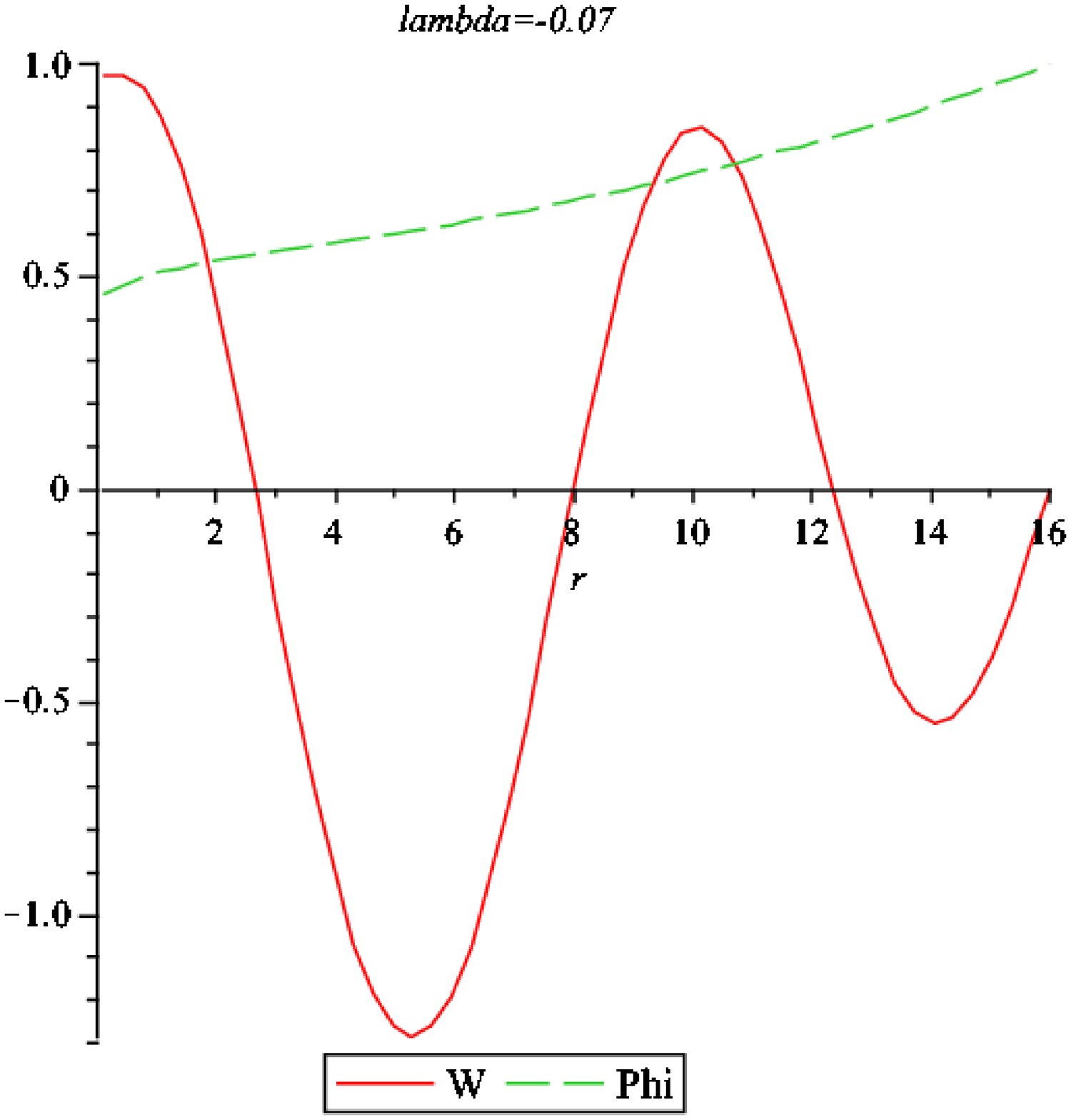}
 \includegraphics[width=0.55\textwidth]{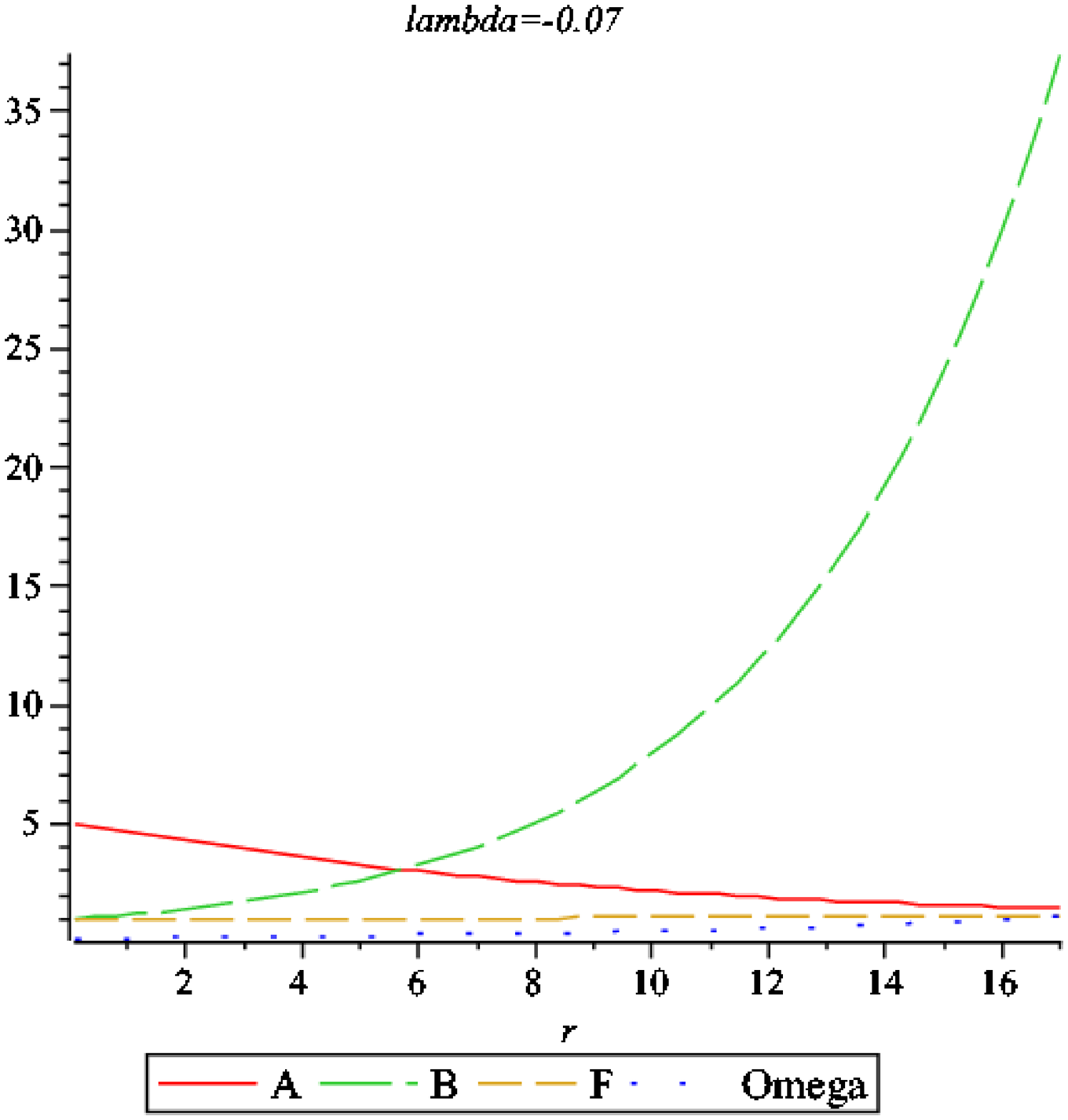}
\hfill}
\centerline{
 \hfill \includegraphics[width=0.55\textwidth]{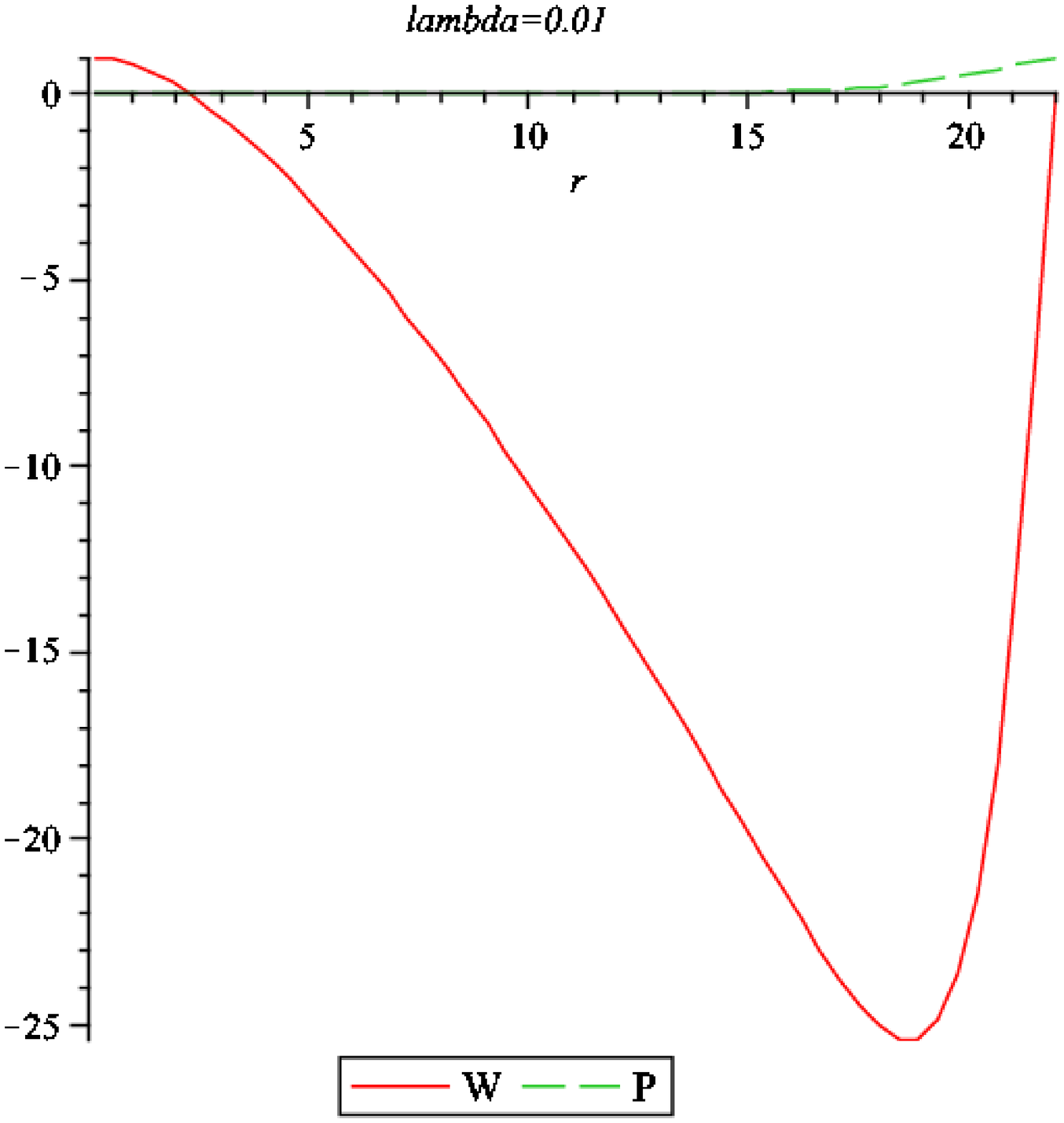}
 \includegraphics[width=0.55\textwidth]{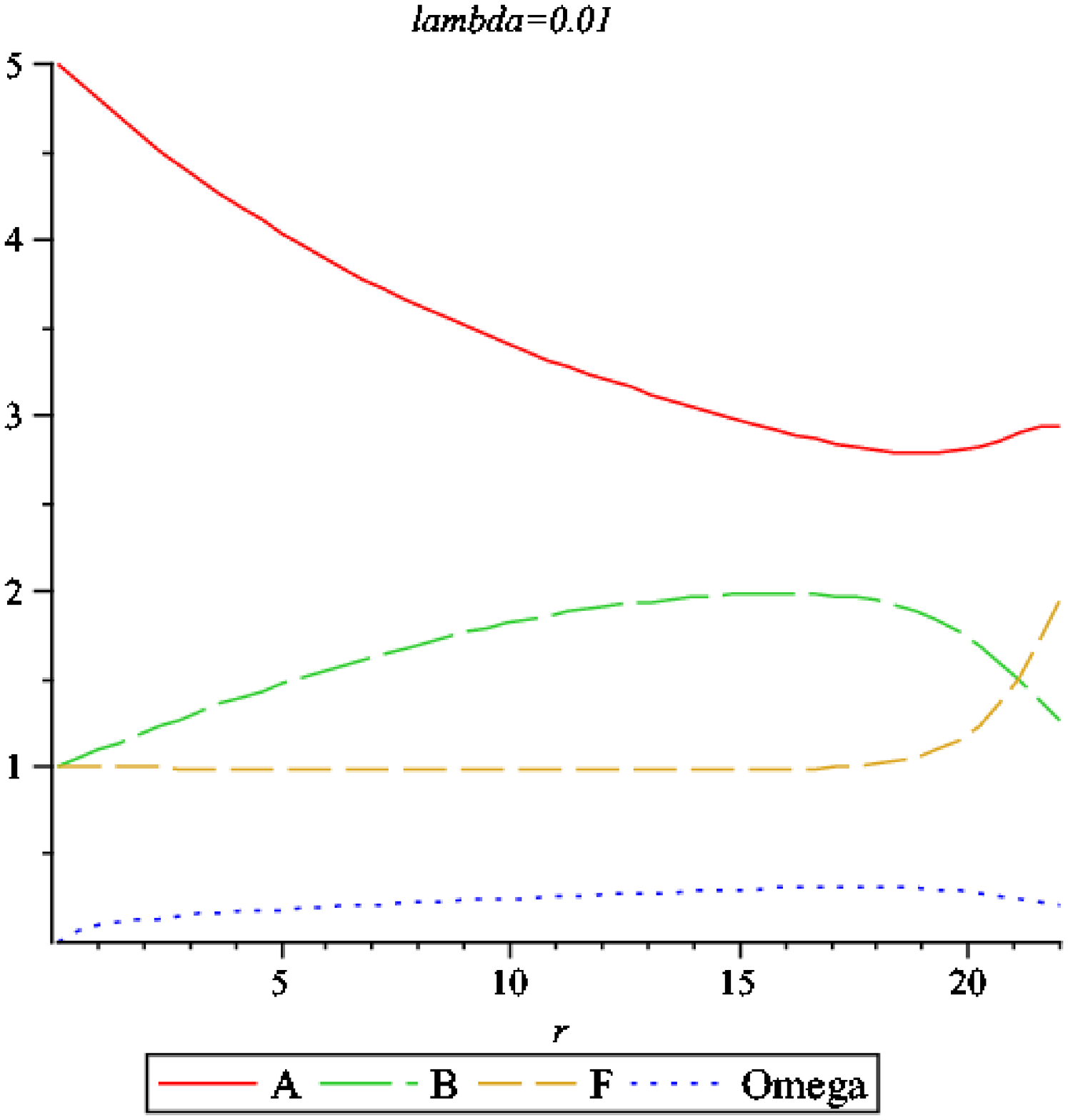}
\hfill}
\caption{Solution with boundary values for the YM components: $W\rightarrow 0, \Phi\rightarrow 1$ at the endpoint of r for $\lambda=-0.07$ and 0.01 respectively. The solution with negative $\Lambda$ shows an acceptable behaviour for $W$.}
\label{fig:3}
\end{figure*}
\begin{figure*}
\centerline{
 \hfill \includegraphics[width=0.55\textwidth]{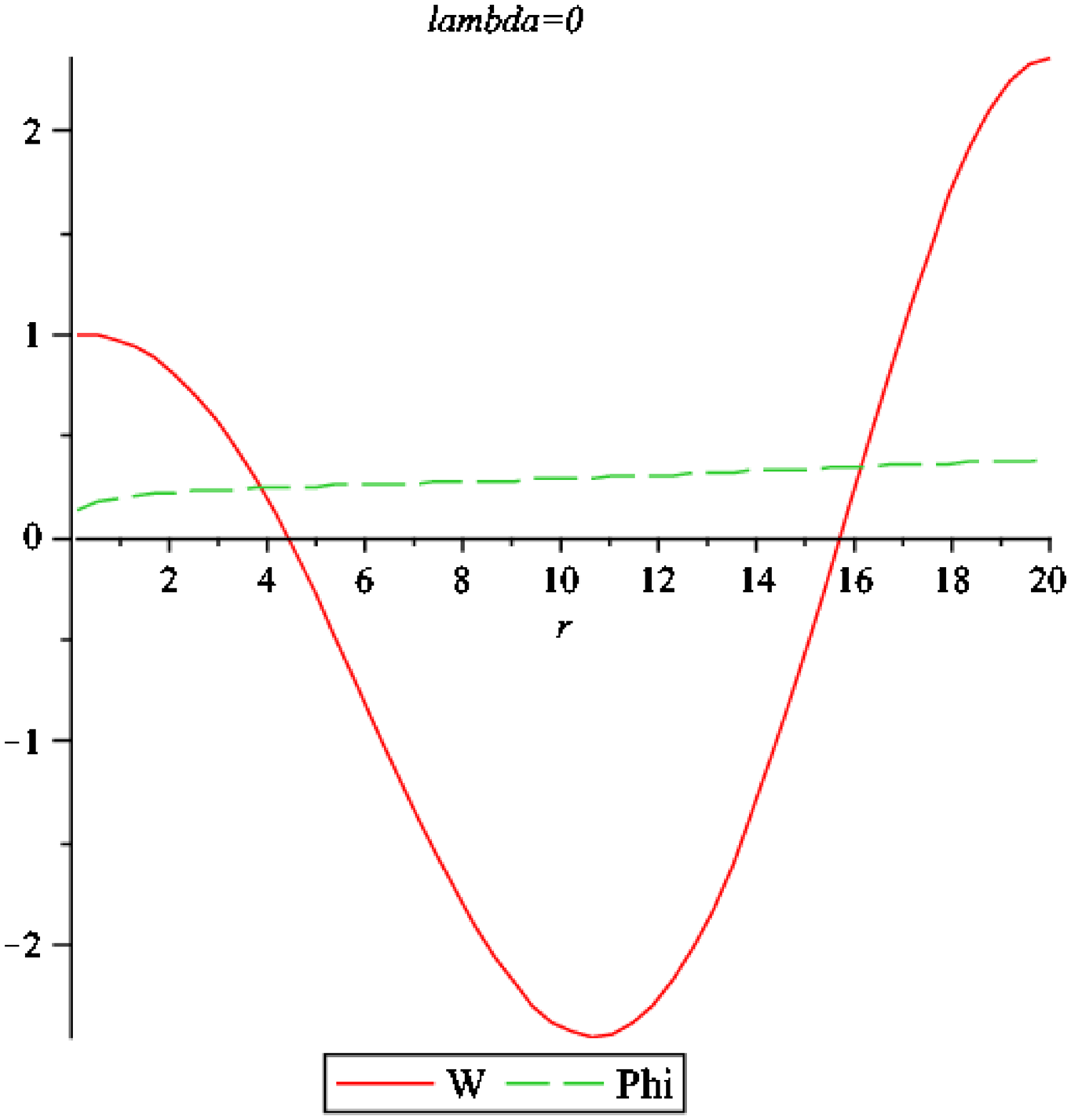}
 \includegraphics[width=0.55\textwidth]{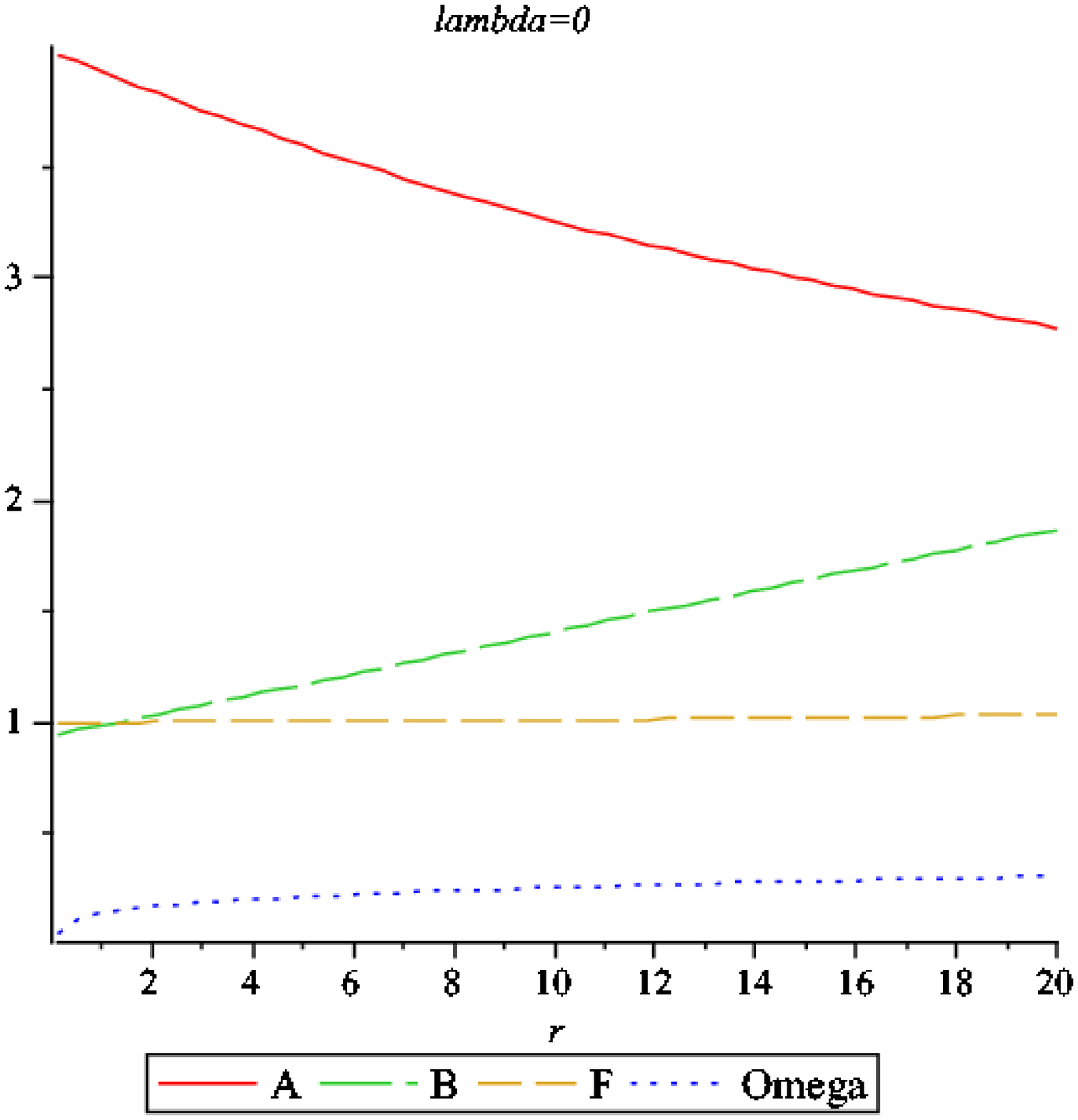}
\hfill}
\centerline{
 \hfill \includegraphics[width=0.55\textwidth]{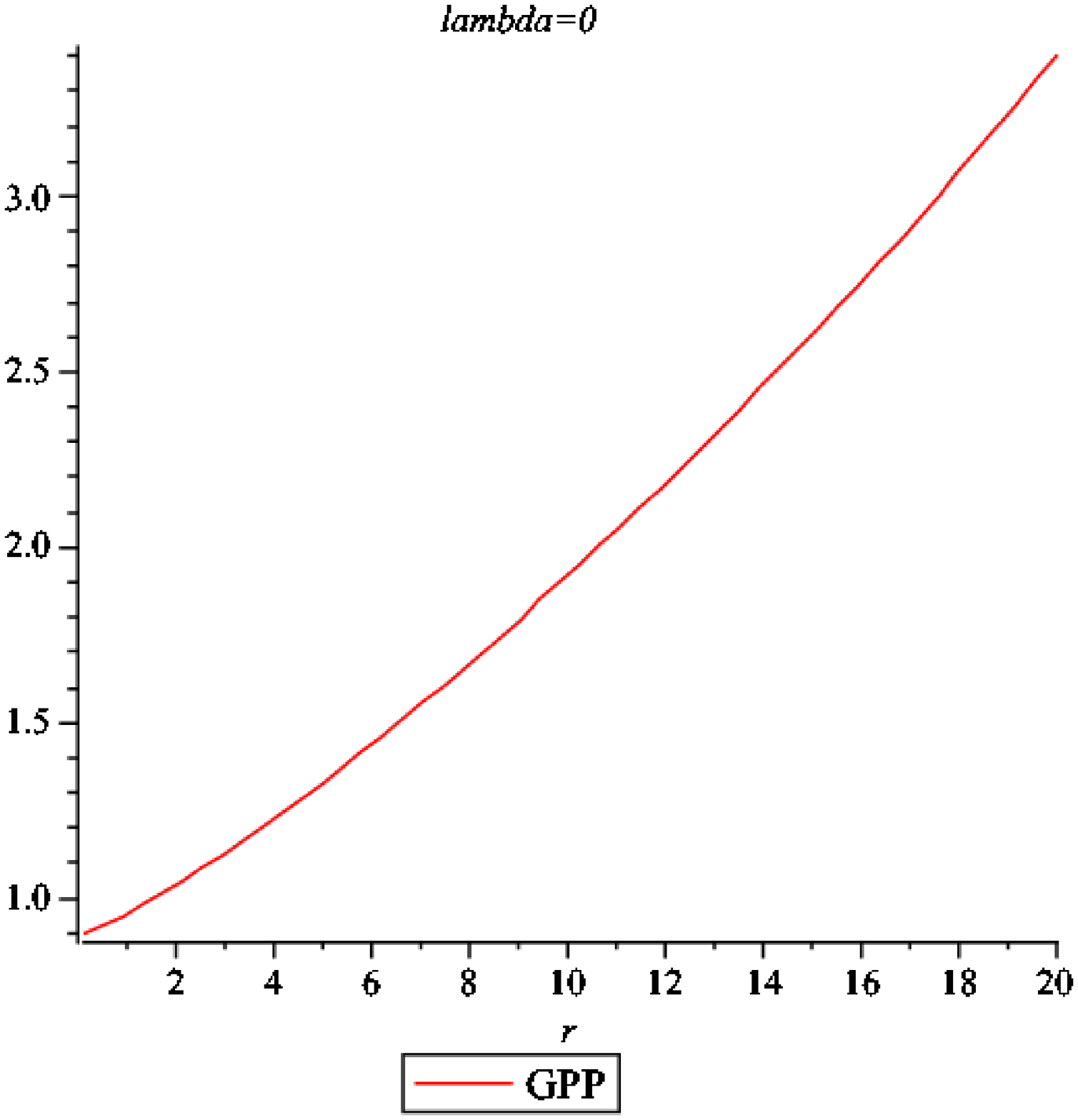}
\hfill}
\caption{$\Lambda=0$ solution with boundary values at the right boundary for the YM components: $W'\rightarrow 0, \Phi'\rightarrow 0$.
The solution is regular with two nodes of $W$.}
\label{fig:4}
\end{figure*}

\begin{figure*}
\centerline{
 \hfill \includegraphics[width=0.55\textwidth]{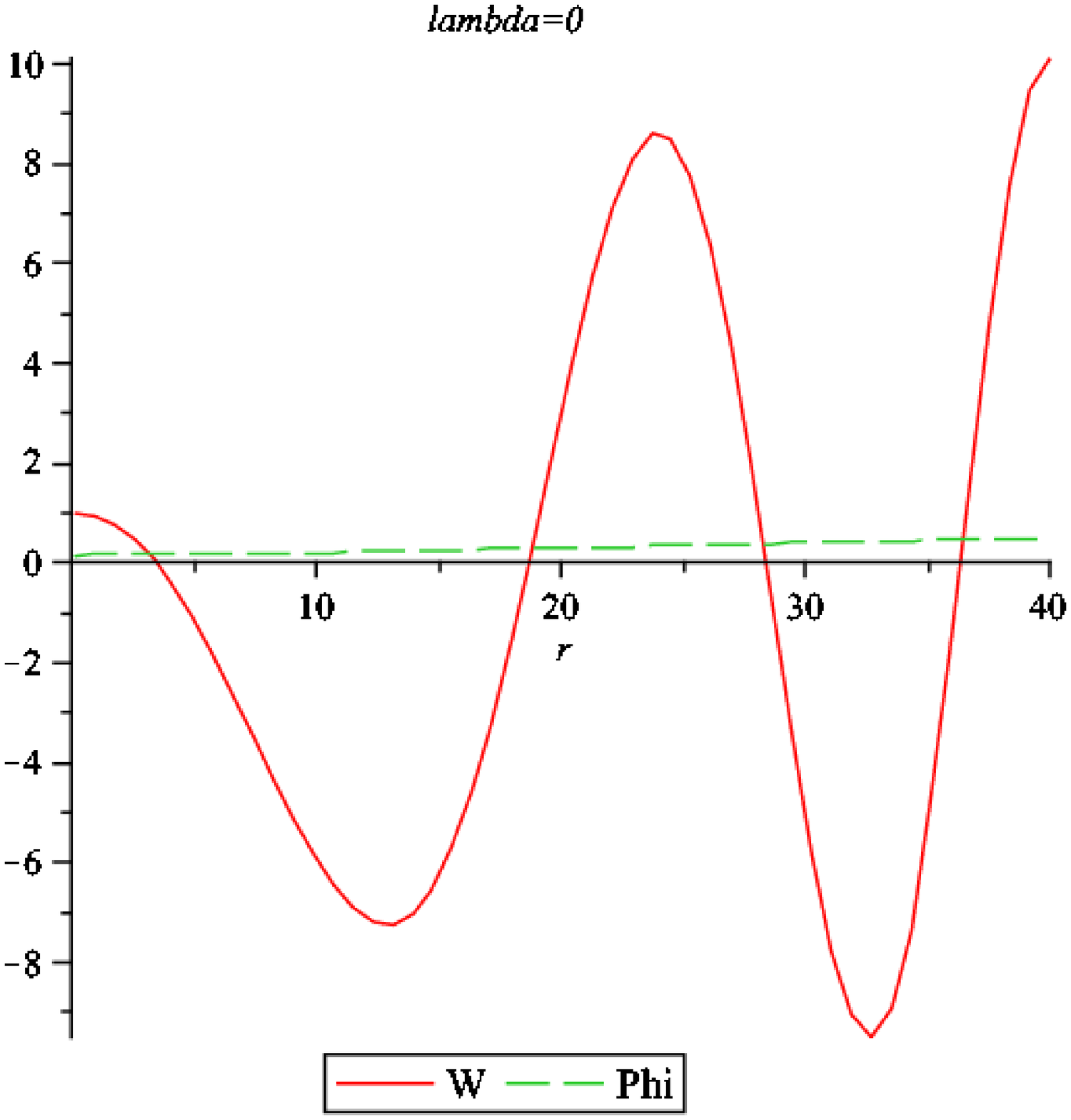}
 \includegraphics[width=0.55\textwidth]{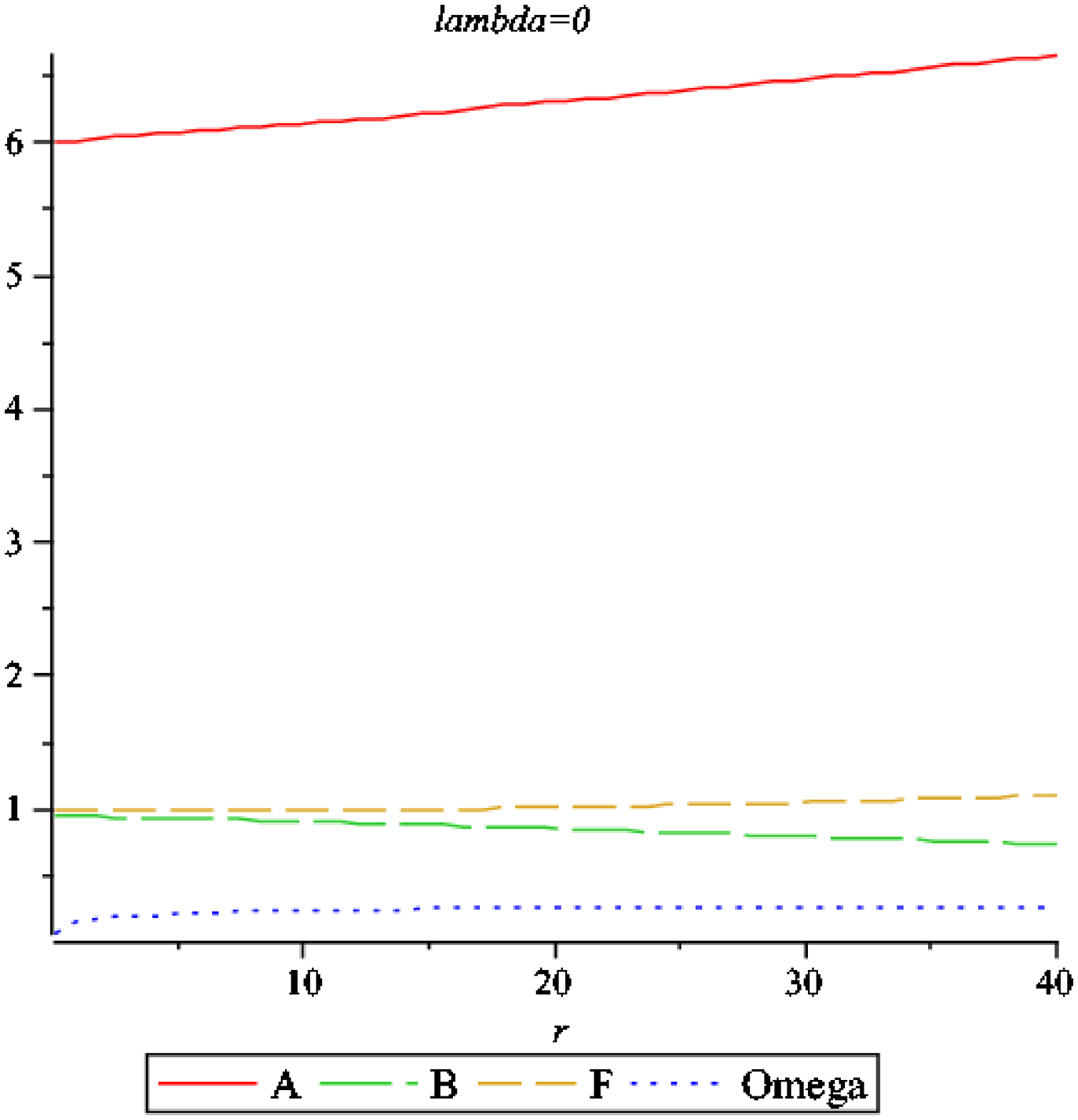}
\hfill}
\centerline{
 \hfill \includegraphics[width=0.55\textwidth]{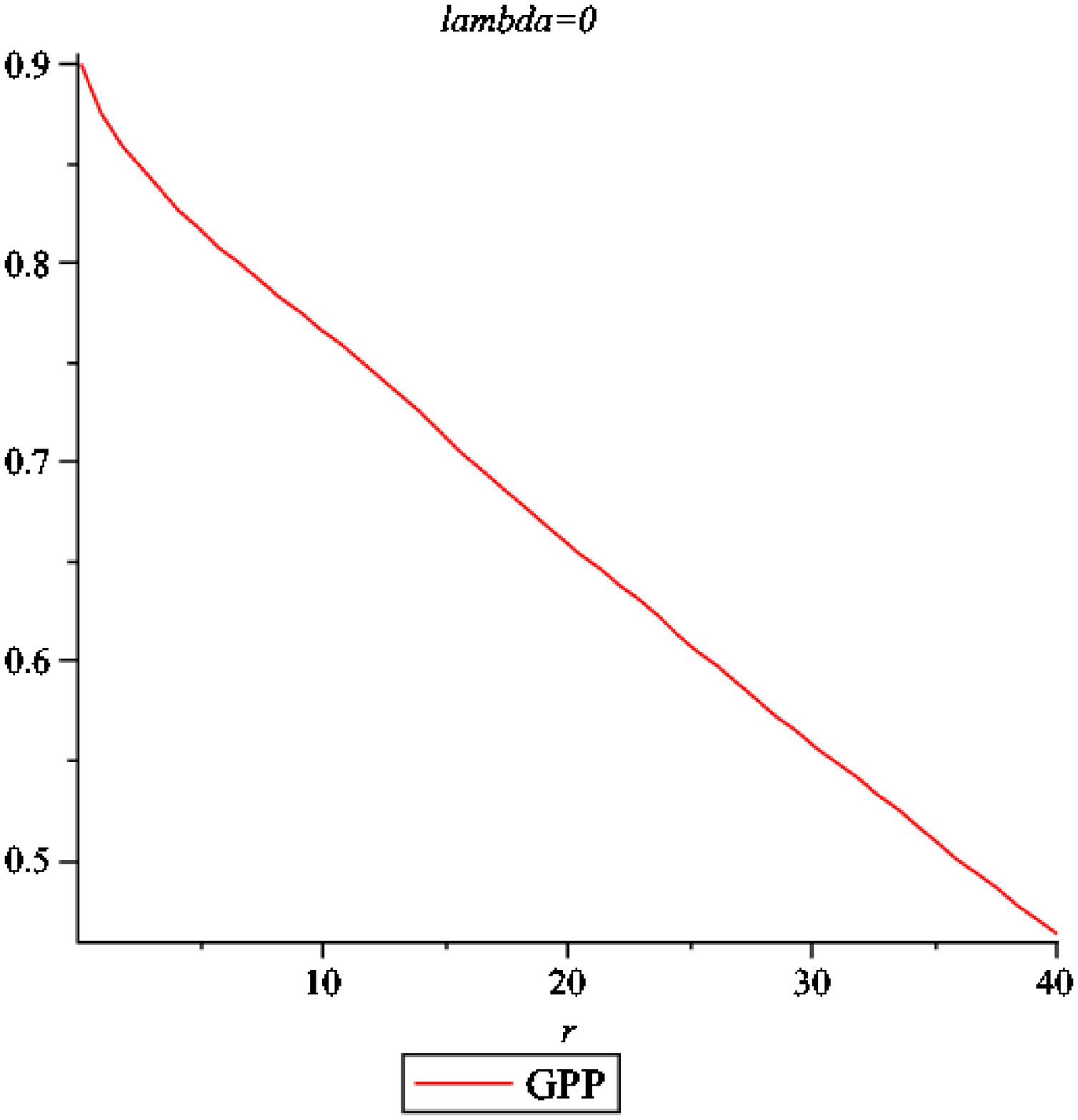}
\hfill}
\caption{As figure 4 but now with a different initial value for $A$. Now $W$ starts oscillating with increasing amplitude and $g_{\psi\psi}$
tends to zero. }
\label{fig:5}
\end{figure*}

\begin{figure*}
\centerline{
 \hfill \includegraphics[width=0.55\textwidth]{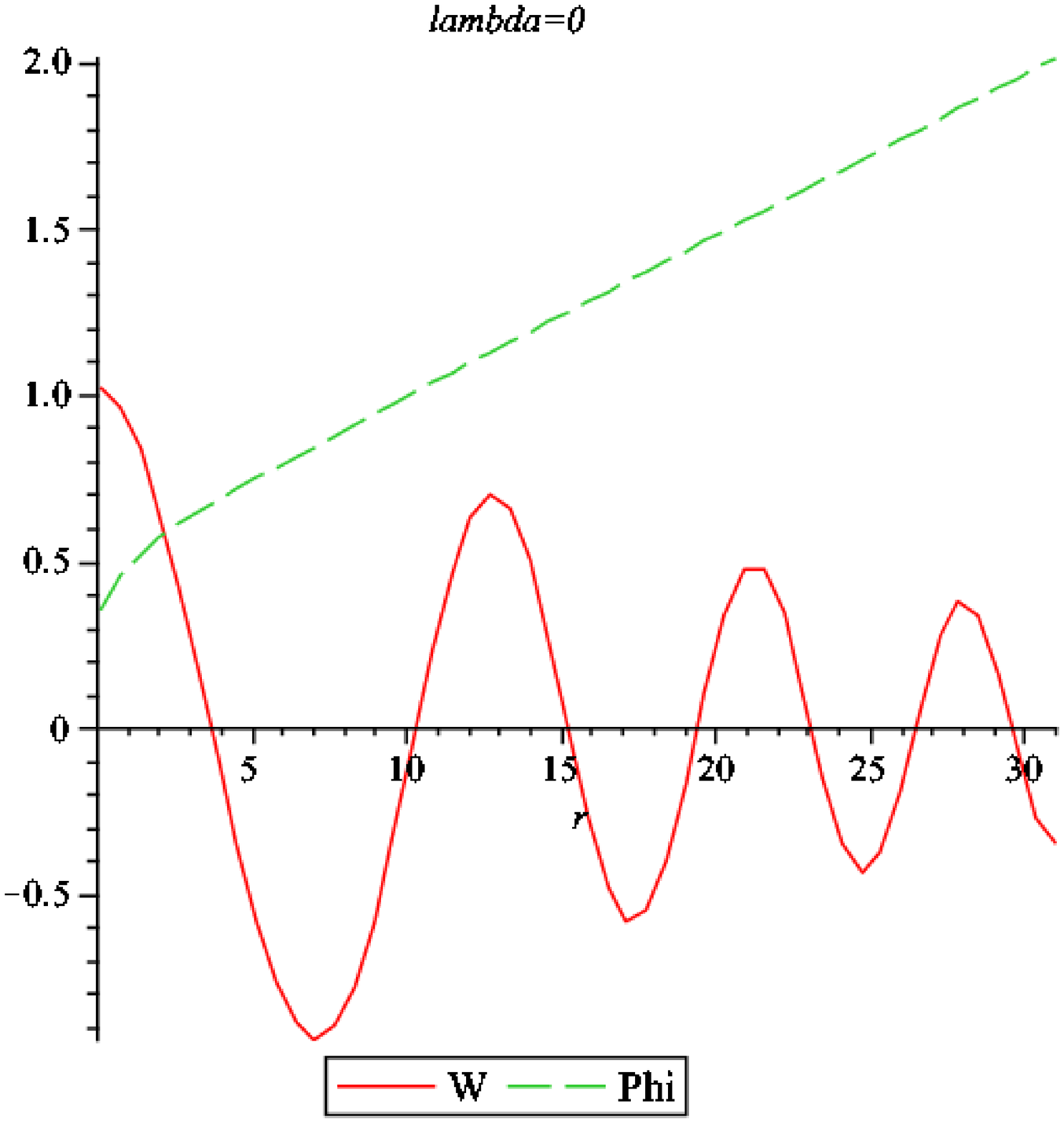}
 \includegraphics[width=0.55\textwidth]{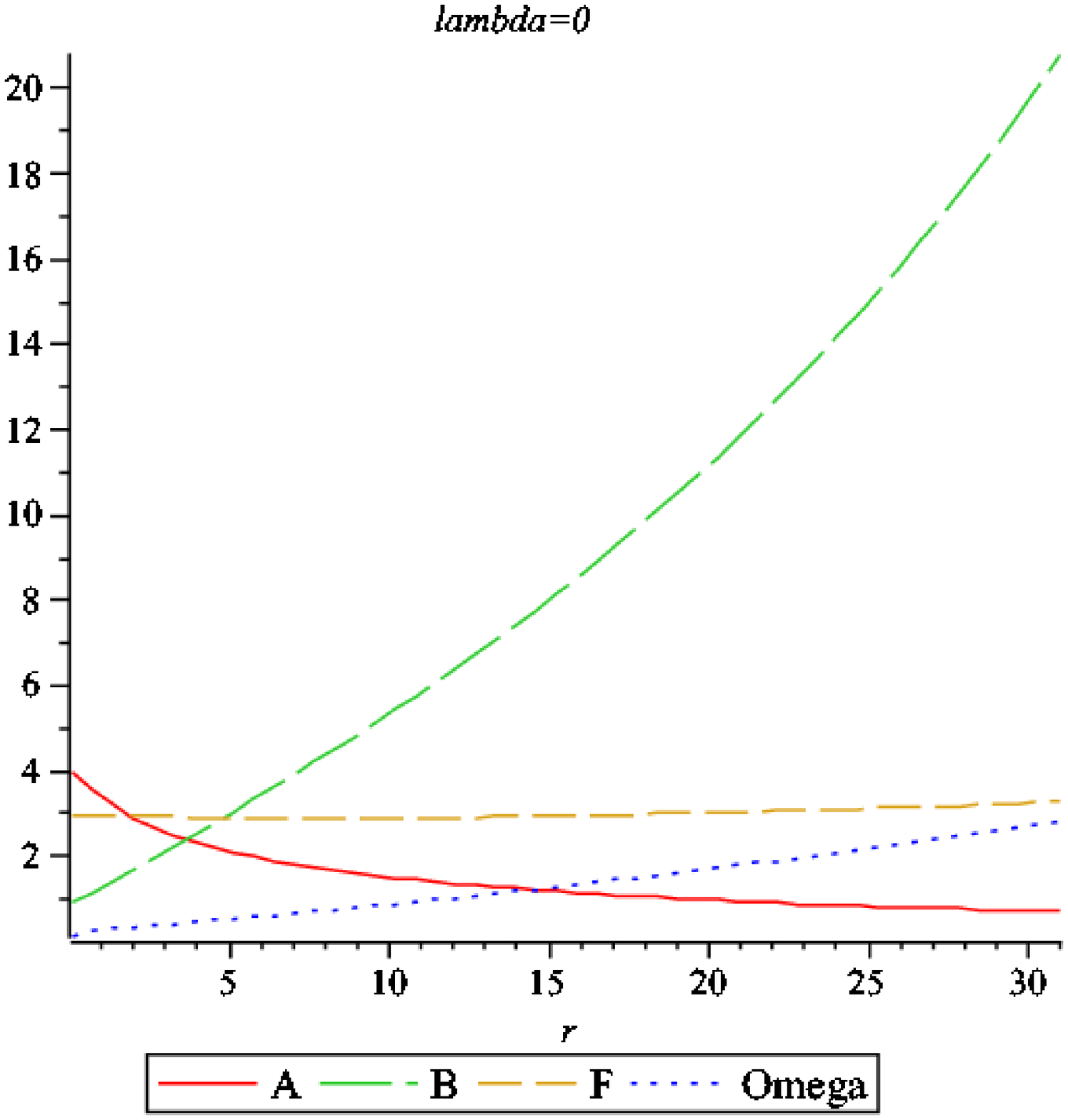}
\hfill}
\centerline{
 \hfill \includegraphics[width=0.55\textwidth]{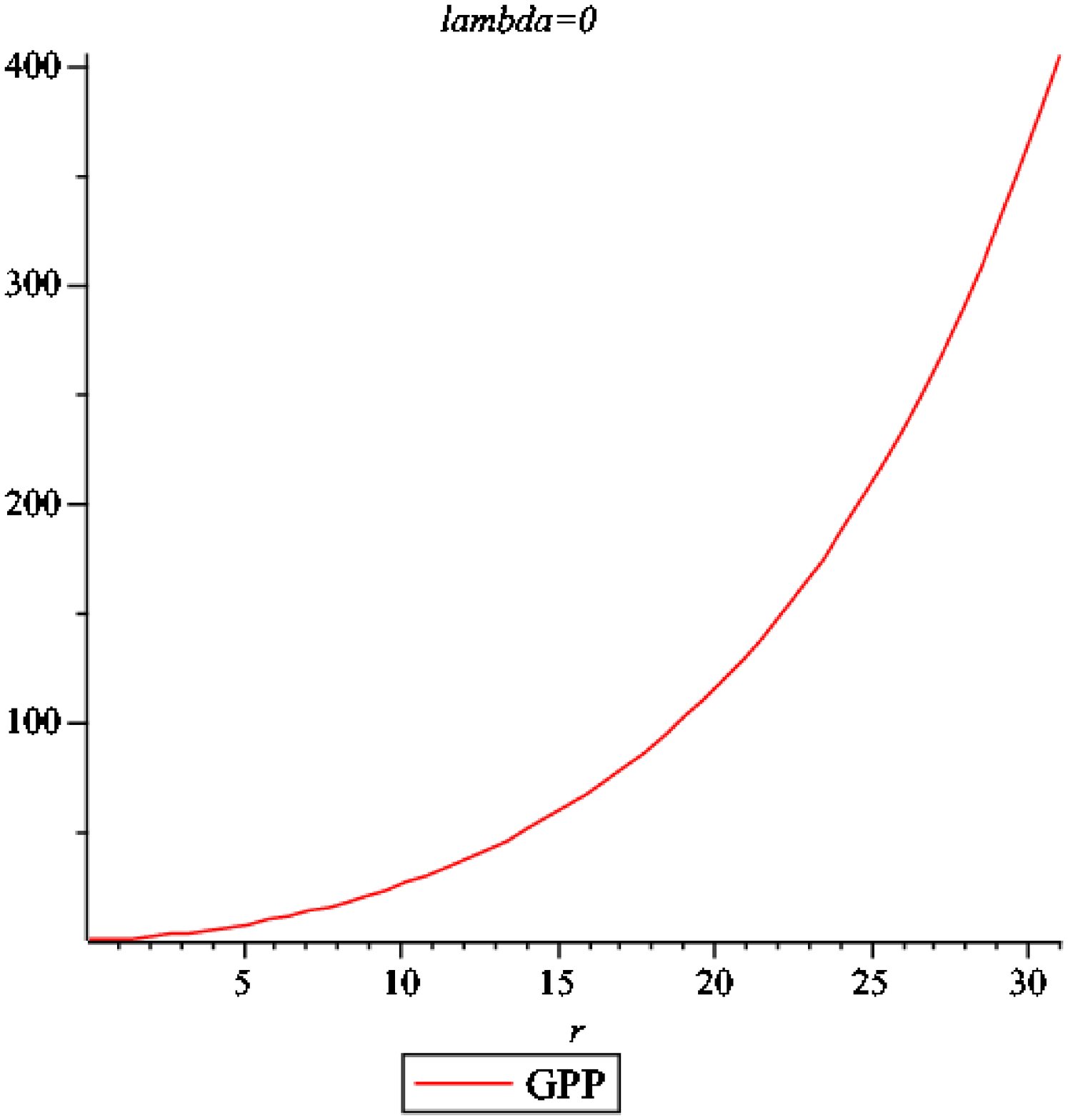}
 \hfill \includegraphics[width=0.55\textwidth]{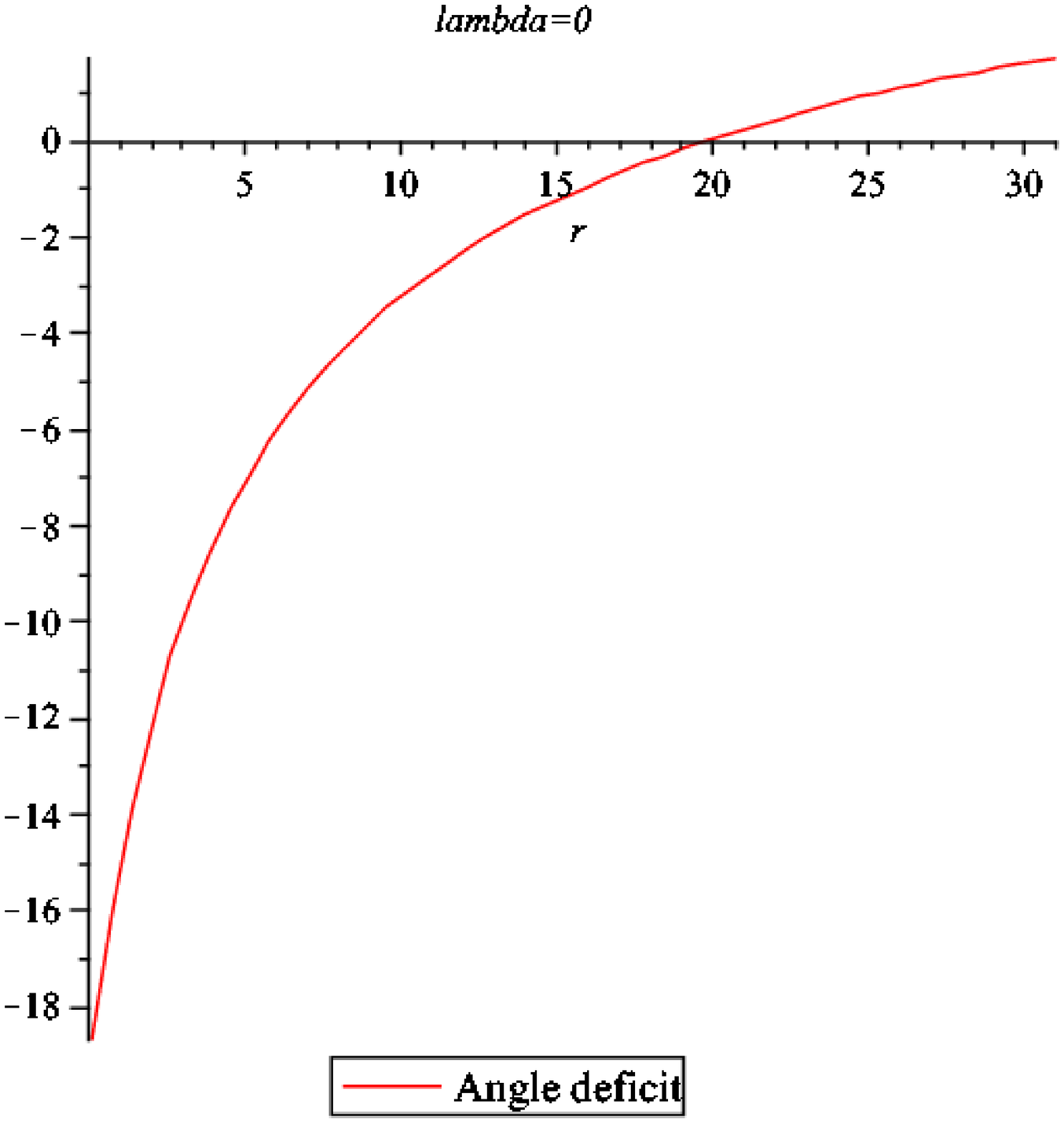}
\hfill}
\caption{$\Lambda=-0.005$ solution with slightly different initial values. Now we have $W$ oscillating around zero with decreasing amplitude.
There is a angle deficit.}
\label{fig:6}
\end{figure*}

\begin{figure*}
\centerline{
 \hfill \includegraphics[width=0.55\textwidth]{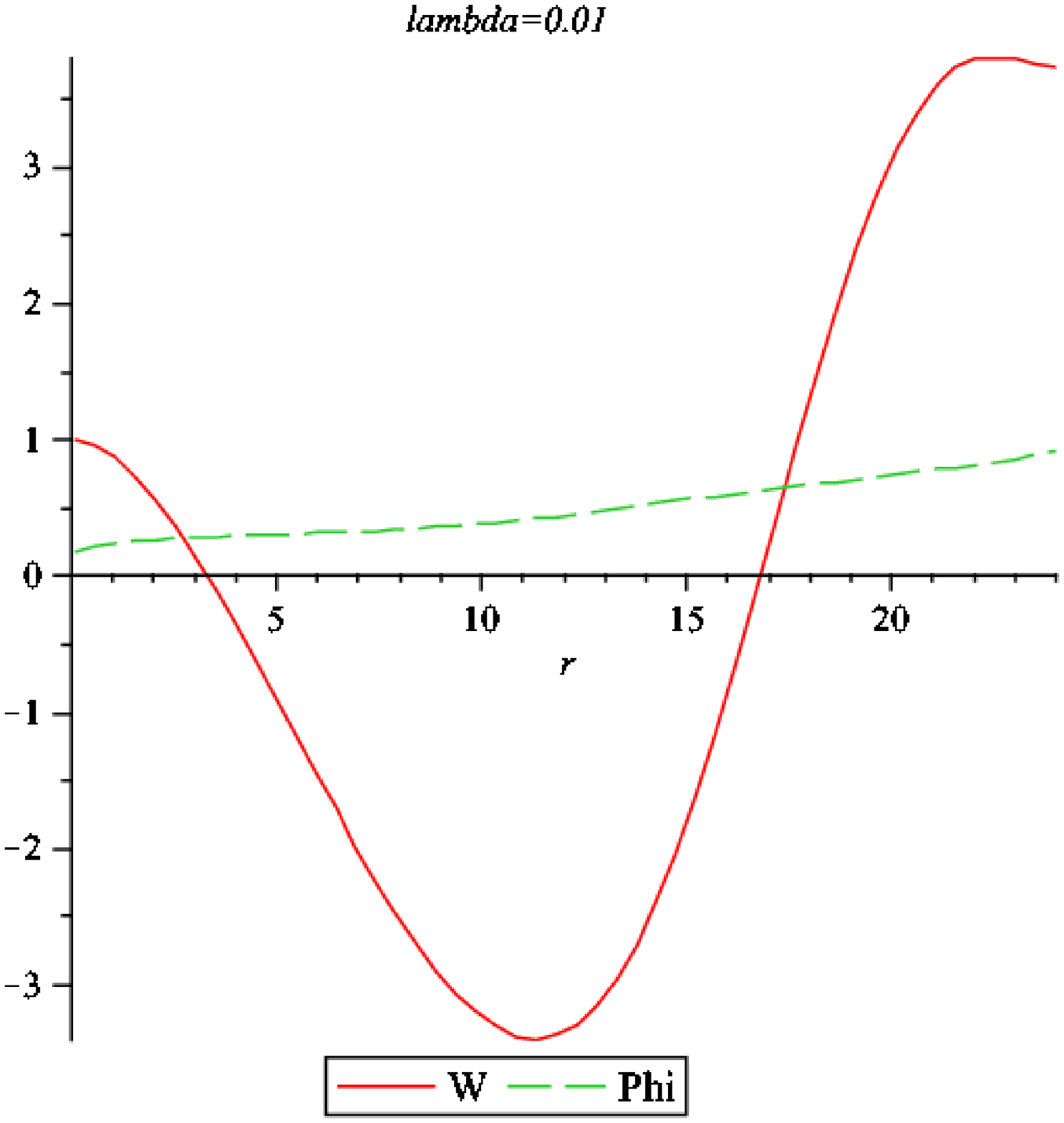}
 \includegraphics[width=0.55\textwidth]{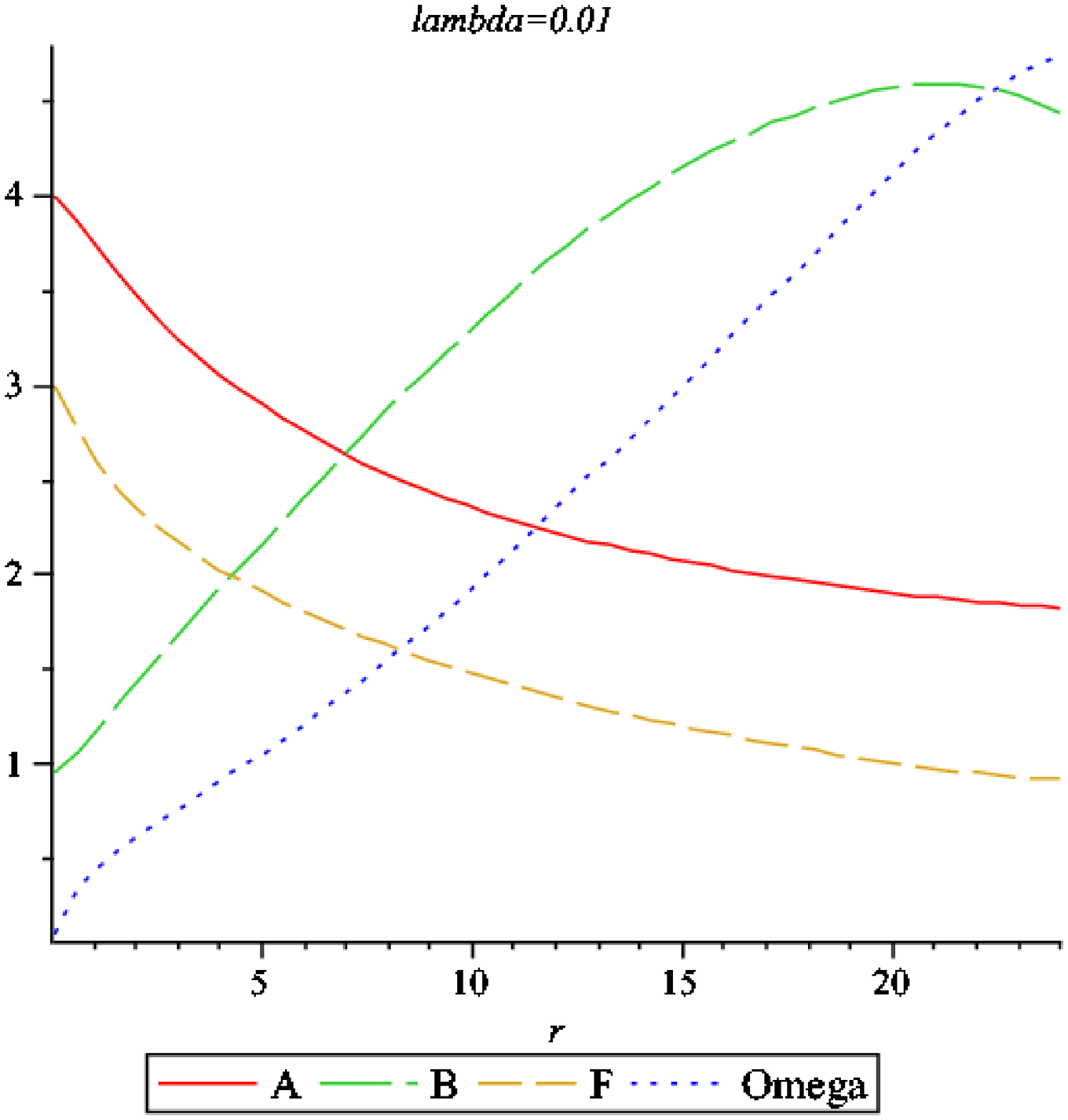}
\hfill}
\centerline{
 \hfill \includegraphics[width=0.55\textwidth]{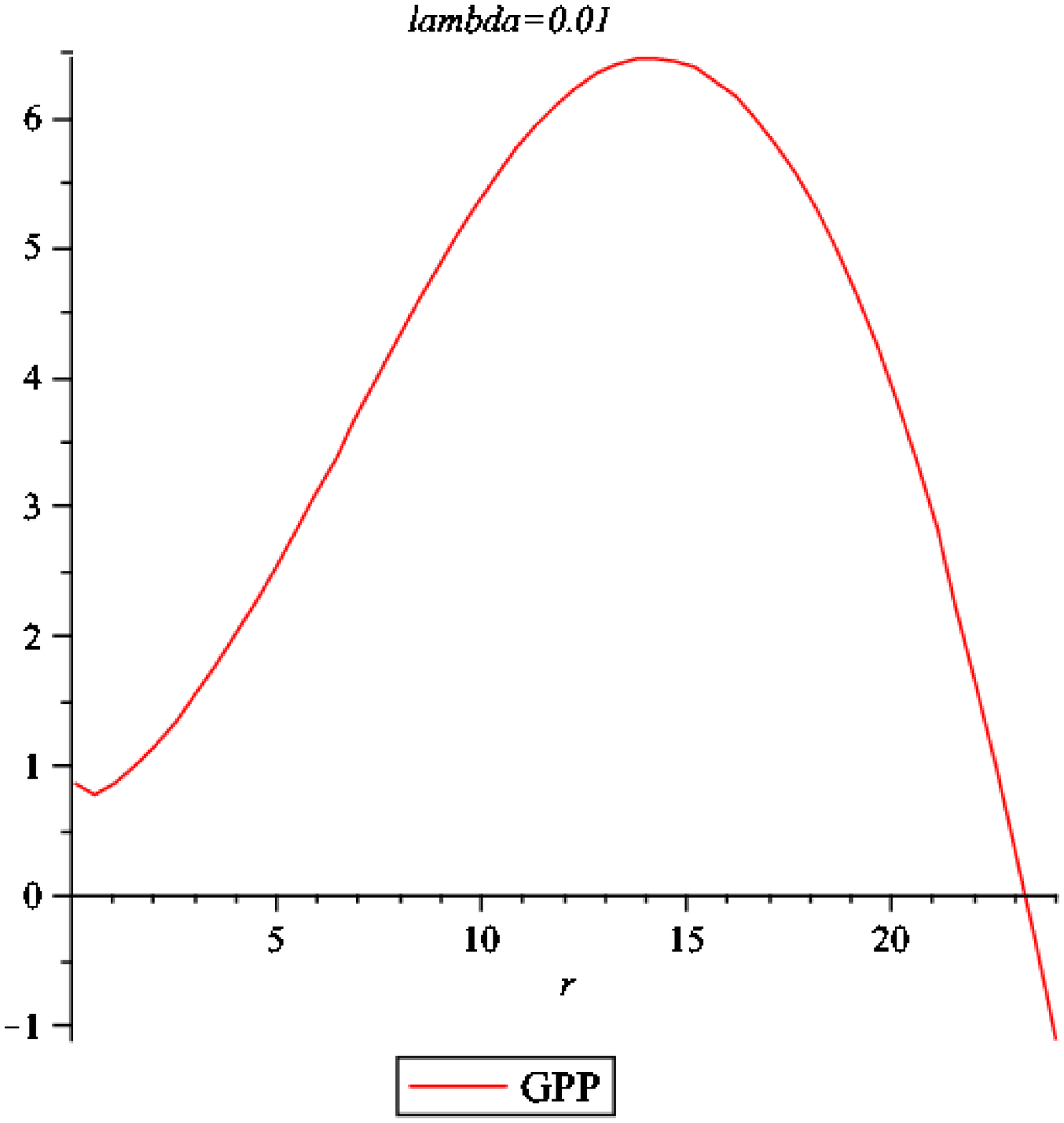}
\hfill}
\caption{$\Lambda=+0.01$ solution. Now a CTC has formed: ($g_{\psi\psi}<0$).}
\label{fig:7}
\end{figure*}

\section{Construction of a Gott space time}

Let us now consider the (3+1) dimensional flat space:
\begin{equation}
ds^2=-dT^2+dx^2+dx'^2+dy^2\label{eq26}
\end{equation}
and apply the transformation to toroidal coordinates $r, \varphi, \psi$ ($0<r<\infty, 0<\varphi <2\pi , 0<\psi<2\pi$),
\begin{eqnarray}
x=\frac{a\sinh r\cos \varphi}{\cosh r-cos\psi},\quad x'=\frac{a\sinh r\sin \varphi}{\cosh r-cos\psi},\quad y=\frac{a\sin \psi}{\cosh r-cos\psi}.
\label{eq27}
\end{eqnarray}
We then obtain the metric
\begin{equation}
ds^2=-dT^2+\frac{a^2}{(\cosh r-\cos\psi)^2}(dr^2+\sinh^2 rd\varphi^2+d\psi^2).\label{eq28}
\end{equation}
If we change further the radial coordinate by $\rho=\frac{\cosh r -1}{\sinh r}$, we have
\begin{equation}
ds^2=-dT^2+\frac{4a^2}{(1+\rho^2-(1-\rho^2)\cos\psi)^2}\Bigl(d\rho^2+\rho^2d\varphi^2+\frac{(1-\rho^2)^2}{4}d\psi^2\Bigr),\label{eq29}
\end{equation}
or,
\begin{equation}
ds^2=-dT^2+\frac{4a^2\cos^2\rho'}{(1+\sin^2\rho'-\cos^2\rho'\cos\psi)^2}\Bigl(d\rho'^2+\tan^2\rho'd\varphi^2+\frac{1}{4}\cos^2\rho'd\psi^2\Bigr),\label{eq30}
\end{equation}
with $\rho'=\arcsin\rho$.
We observe that $\rho\rightarrow 1$ when $r\rightarrow \infty$. This metric is almost the conformal static analogue of (Eq.13) in (3+1)dimensional
sub-space by skipping the $dz^2$ term, with two different angle deficits in $\varphi$ and $\psi$.
We can perform the following matching condition
\begin{equation}
{\cal R}=L_x\Omega_{xy}\Omega_{xx'}\Omega_{x'y}L_x,\label{eq31}
\end{equation}
with $\Omega_{xy}, \Omega_{x'y}$ rotations in the $(x-y)$ and $(x'-y)$-planes, and $L_x$ a boost:

\begin{eqnarray}
    \Omega_{xy}=\left[\begin{array}{cccc}
              \cos 2\beta_1&\sin 2\beta_1&0&0\\-\sin 2\beta_1&\cos 2\beta_1&0&0\\0&0&1&0\\0&0&0&1
               \end{array}\right]
             \qquad  \Omega_{x'y}=\left[\begin{array}{cccc}
             \cos 2\beta_2&0&\sin 2\beta_2&0\\0&1&0&0\\-\sin 2\beta_2&0&\cos 2\beta_2&0\\0&0&0&1
                \end{array}\right]\nonumber \label{eq32}
 \end{eqnarray}

\begin{eqnarray}
    \Omega_{xx'}=\left[\begin{array}{cccc}
              1&0&0&0\\0&0&1&0\\0&-1&0&0\\0&0&0&1
               \end{array}\right]
             \qquad  L_x=\left[\begin{array}{cccc}
             \cosh \gamma&0&0&\sinh \gamma\\0&1&0&0\\0&0&1&0\\\sinh\gamma&0&0&\cosh\gamma
                \end{array}\right],\label{eq33}
 \end{eqnarray}
with $\beta =4\pi G m =\pi(1-\alpha), \alpha $ the angle deficit and m the mass.
We obtain for the trace of ${\cal R}$
\begin{equation}
{\bf Tr}({\cal R})=1-\sin 2\beta_1\sin 2\beta_2+\cos 2\beta_2\cos 2\beta_2.\label{eq34}
\end{equation}
For $\beta_1=\beta_2=\beta$ we obtain
\begin{equation}
{\bf Tr}({\cal R})=2\cos^2 2\beta ,\label{eq35}
\end{equation}
with $\cos 2\beta <1$.
 Let us now consider 2 opposite moving 5-D strings. The matching condition then yields:
\begin{equation}
{\bf Tr}{\cal R}=4(\cos^22\beta -1)^2(1-2\cosh^2\xi)^2.\label{eq36}
\end{equation}
From the effective $\Omega_{eff}$ we obtain then the inequality
\begin{equation}
(\cos^22\beta -1)(1-2\cosh^2\xi)<\frac{1}{2}\sqrt{3}.\label{eq37}
\end{equation}

Let us now consider the Gott condition $\cosh\xi >\frac{1}{\sin\beta}$. If we choose the minimal possibility, we
obtain then  the condition on $\cosh\xi$:
\begin{equation}
1<\cosh\xi<\frac{\sqrt{110+22\sqrt{3}}}{11}\approx 1.11 \label{eq38}
\end{equation}
So there could be, in principle, a CTC. Now it is believed that in (3+1)dimensional space times which have
physically acceptable global structure, CTC's will not occur. This was outlined in section 2. In our (4+1) dimensional
global space time there could be a situation where Gott's condition is fulfilled. This is only true for  non-interacting
strings using the "glue and paste" approach of section 2.

\section{ Conclusion and outlook}

It is known that the spacetime of a spinning cosmic string is endowed with an unusual
topology and could generate the controversial closed timelike curves. The
increase in interest in these models originates not only from the fact that
causality violation could occur, but also from the conjecture that the
solution of these controversies could be related to a possible quantum
version of such systems \cite{Ana}.
Although one might prove that the evolution of CTC's
can be prevented in our universe \cite{Des2,Hooft}, the dynamically formed topology changes in
some non-vacuum systems still remain intriguing \cite{Slag3,Slag5}.

Here we investigated the cosmic string-like features of in the EYM model in 5-dimensional spacetime. Where in the 4-dimensional
case it was proved that the effective two-particle generator of the isometry group becomes hyperbolic (spacelike), i.e., contradicting
Gott's condition, in the 5-dimensional case there could exist a Gott spacetime. This holds only if we may freely glue together the
two topologies of the moving strings, just as in the 4-dimensional case.

We also presented numerical solutions of the complete system of equations. It is conjectured 
that CTC's will arise only with singular behaviour of the metric. Moreover, a negative cosmological constant would improve
the regularity of the solutions. If we impose the usual boundary values for the YM components far from the string, i.e., 
$W\rightarrow 0$ and $\Phi\rightarrow 1$, then again a negative $\Lambda$ improves regularity.
Moreover we find an angle deficit in our model. 

Vortex like  behaviour in higher dimensional gravity models with  non-Abelian gauge fields is not uncommon\cite{Nak}. One believes
that Abelian gauge symmetry, which is used in the "conventional" cosmic string models, might come from non-Abelian symmetry, where the U(1)
symmetry is embedded in the higher symmetry. A cosmic string-like solutions in these models is then obtained by dimensional reduction 
and a relation
between $\Lambda, g$ and $ G_5$. The angle deficit and the size of the extra dimension then depends on this relation.
We find similar behaviour. It is quite surprising that cosmic string-like behaviour is found without the Higgs potential and
specific value of the vacuum expectation value of the Higgs field. 

The next step must be the dynamical investigation of our model. This is currently under study by the author.

\section*{Acknowledgments}
I am very gratefully to Prof. Dr. D. H. Tchrakian and Dr. E. Radu for the hospitality and useful discussions during my visit
at Maynooth National University of Ireland.


\section*{References}
\begin{thebibliography}{25}
\bibitem{Ran}Randall L and Sundrum R, 1999  {\it Phys. Rev. Lett.} {\bf 83}, 3370 
\bibitem{Vol}Volkov M and Gal'tsov D, 1999  {\it Phys. Rep.} {\bf 319} 1
\bibitem{Mal}Maldcena J, 1999  {\it Adv. Theor. Math. Phys.} {\bf 2} 3370, 231 
\bibitem{Hos}Bjoraker J and Hosotani Y, 2000 {\it Phys. Rev. Lett.} {\bf 84} 1853 
\bibitem{Vil}Vilenkin A and  Shellard E, 1994   {\it Cosmic Strings and Other Topological Defects} (Cambridge University Press,
Cambridge)
\bibitem{Mar}Marder L, {\it Proc. Roy. Soc.} A{\bf 252} 45 
\bibitem{Gott}Gott J R, 1990  {\it Phys. Rev. Lett.} {\bf 66} 1126 
\bibitem{Des1}Deser S, Jackiw R and  't Hooft G, 1983  {\it Ann. of Phys.} {\bf 152} 220 
\bibitem{Des2} Deser S, Jackiw R and  't Hooft G, 1992 {\it Phys. Rev. Lett.} {\bf 68} 267 
\bibitem{Slag1} Slagter R J, 1997 in {\it Proc. of 8th Marcel Grossmann Meeting, Jerusalem}, eds. T. Piran and R. Ruffini
(World Scientific, Singapore), 602
\bibitem{Slag2} Slagter R J, 1996 {\it Phys. Rev.} D{\bf 54} 4873 
\bibitem{Hol} Holst S J, 1995   {\it Preprint} gr-qc/9501010
\bibitem{Bir} Birmingham D and   Siddhartha S, 1999 {\it Phys. Rev. Lett.} {\bf 84} 1074 
\bibitem{Slag3} Slagter R J, 2005 to appear in {\it  Proc. of the Conference on General Relativity and Gravitation, Paris}
\bibitem{Slag5} Slagter R J, 2002 {\it Class. Quantum Grav.} {\bf 19} 115 
\bibitem{Car} Carrol S M,  Fahri E and Guth A H, 1992   {\it Phys. Rev. Lett.} {\bf 68} 263 
\bibitem{Hooft}'t Hooft G, 1992 {\it Class. Quantum Grav.} {\bf 9} 1335 
\bibitem{Dy}Dyer C and Marleau R, 1995 {\it Phys. Rev.} D {\bf 52} 5588 
\bibitem{Lag}Laguna P and Garfinkle D, 1989 {\it Phys. Rev.} D {\bf 40} 1011 
\bibitem{Laguna}Laguna-Castillo P and Matzner R A, 1987 {\it Phys. Rev.} D {\bf 36} 3663 
\bibitem{Vol2}Volkov M S, 2001 {\it Preprint} hep-th/0103038
\bibitem{Okuy}Okuyama N and Maeda K, 2002 {\it Preprint} gr-qc/0212022
\bibitem{Nak}Nakamura A and  Hirenzaki S, 1990 {\it  Nucl. Phys.} B {\bf 339}, 533 
\bibitem{Ana}Anadan J, 1996 {\it Phys. Rev.} D{\bf 53}, 779 


\end{thebibliography}
\end{document}